# Super-resolution Imaging of Limited-size Objects

Taeyong Chang[1]*, Giorgio Adamo[1,2]*, Nikolay I. Zheludev[1,2,3]

[1] Centre for Disruptive Photonic Technologies, SPMS and TPI, Nanyang Technological University, Singapore 637371, Singapore

[2] Optoelectronics Research Centre and Centre for Photonic Metamaterials, University of Southampton, Southampton SO17 1BJ, UK

[3] Hagler Institute for Advanced Study, Texas A&M University, Texas 77843, USA

*Correspondence to: taeyong.chang@ntu.edu.sg, g.adamo@soton.ac.uk.

## Abstract

Improvement of label-free far-field resolution of optical imaging is possible with prior knowledge of the object such as its sparsity or accumulated by *a posteriori* examination of a similar class of object[1-4]. We show that the sole knowledge of the object's limited size is another fundamental resource to achieve resolution beyond the Abbe-Rayleigh diffraction limit: a higher resolution can be achieved with smaller objects. To prove this, we developed an imaging method that involves the representation of light scattered from the object with orthonormal field-of-view-limited Slepian-Pollak functions and experimentally demonstrated $\lambda/8$ resolution of sub-wavelength objects. Our method requires no assumption of the shape and complexity of the object and its labelling allowing a wide range of applications in the studies of nanoparticles and isolated microorganisms.

## Introduction

The resolution limit for optical imaging of about half of the wavelength of light ($\lambda/2$), introduced by Abbe[5], Helmholtz[6], and Rayleigh[7] more than a century ago, is generally accepted until today. Its popularity as a resolution criterion does not originate from it being a rigorous or fundamental quantity, but because it applies well to practical microscopy. It is hard to overcome without using sample-invasive techniques, such as near-field scanning and fluorescent staining[8,9]. The possibility of achieving super-resolution in a label-free and far-field configuration can be quantitatively investigated using quantum information theory[1-3], where the imaging system is equivalent to the communication channel, and the imaging performance corresponds to the amount of information reliably transferred from object to observer during the measurement and used to construct the image. While this approach does not provide a universal and simple criterion (like $\lambda/2$), it establishes that super-resolution imaging is attainable given a certain degree of prior knowledge of the object and sufficient measurement accuracy[1-4,10-12].

An increase in imaging resolution by assuming the size limitation of the object under observation has been a long-standing topic of theoretical research efforts, between 1950s and 2000s[13-24], which mathematically established that any resolution is, in principle, achievable. One route to achieve it is by using Slepian-Pollak functions, i.e. prolate spheroidal wave functions[25,26]. Expressing any arbitrary confined function as a superposition of Slepian-Pollak functions is possible, whose coefficients can be uniquely determined by diffraction-limited measurements[17-19]. However, all theoretical investigations agreed that achieving meaningful super-resolution in experiments might not be feasible because of the required high measurement accuracy[4,10-12]. That is why, to our knowledge, no experimental demonstrations of deep super-resolution imaging where the sole assumption is a limited object size have been achieved to date.

Here, we experimentally achieve resolutions of $\lambda/7$ and $\lambda/8$ for, respectively, 2-dimensional (2D) and 1-dimensional (1D) label-free far-field optical imaging of subwavelength objects (<$0.8\lambda$), with coherent visible light ($\lambda$=638 nm) and no assumptions other than the limited size of the object. We first show how our imaging can be understood as a reverse process of superoscillatory hotspot generation – the formation of a free-space electromagnetic field profile that locally oscillates faster than its highest frequency component. Both can be mathematically expressed with Slepian-Pollak functions, and this allows us to identify the experimental criticalities shared by the two processes. Then, we reformulate the theory of Slepian-Pollak imaging within an information framework to quantify the fundamental bound of the imaging performance and show the trade-off relationship between resolution, photon budget, and the object size limit. In the experiment, we perform repeated shot-noise limited measurements to obtain sufficient information about the imaged objects and apply a vector linear filter to correct the distortion present in the measurement setup. The determination of the required number of measurement repetitions to retrieve each Slepian-Pollak coefficient and the development of a suitable vector linear filter, based on information theory, represent breakthroughs that allowed us to achieve super-resolution. We experimentally demonstrate point spread functions (PSFs) much narrower than that of an ideal conventional free-space imaging system, and successfully image 2D and 1D nanoscale objects of various shapes.

## Results

**The analogy between superoscillatory hotspot generation and Slepian-Pollak imaging**

A simple understanding of the Slepian-Pollak imaging can be derived if one recognizes it as a reverse process of superoscillatory hotspot generation (Fig. 1). It is known that an arbitrary wave profile – including a superoscillatory one with arbitrary small spot size – can be created within the field of view (FOV) $D$ on the object plane ($x$) by realizing an appropriate optical source within a band-limited region ($K$) on the Fourier plane ($x'$), where the spatial coordinates are proportional to the spatial frequency coordinates of the wave, as shown in Fig. 1a[27-30]. While such a source can be realized by designer masks with conventional illumination, it has been extensively studied how finding an appropriate design of a mask that takes into account the realistic distortion of the optical setup is challenging, especially in the case of a superoscillatory profile with a large FOV and a narrow hotspot[28-30]. Conversely, in the Slepian-Pollak imaging (Fig. 1b), the optical source (originating from the object) is confined within a region ($D$) on the object plane ($x$), while the measurable wave is band-limited within a region ($K$) on the Fourier plane ($x'$). It is apparent that the Slepian-Pollak imaging and the superoscillatory hotspot generation are reciprocal configurations (see Supplementary Text S1) [31].

In both cases, Slepian-Pollak functions concisely represent the source-wave relationship. In the case of the Slepian-Pollak imaging (superoscillatory hotspot generation), this implies that the $i$th Slepian-Pollak source profile, $S_i$, at the confined object (Fourier) plane, propagates and generates only an $i$th Slepian-Pollak wave profile at the confined Fourier (object) plane, (albeit allowing non-zero wave amplitude outside the confined region). It can be shown that Slepian-Pollak imaging, which amount to finding the coefficients, $c_i$, of the Slepian-Pollak series, $\sum_i c_i S_i(x)$, is straightforward in-principle. If $\sum_i d_i S_i(x')$ is the measured Slepian-Pollak series of the wave profile in a confined region at the Fourier plane, $c_i$ is determined by $c_i = d_i/\gamma_i$ for all $i$ where $\gamma_i$ is the complex amplitude transfer ratio of $i$th Slepian-Pollak function (see Supplementary Text S1) [17-19]. This denotes the in-principle achievability of arbitrarily fine super-resolution imaging, because the solution approximates the source profile with arbitrary accuracy, for large enough $i$. The experimental difficulty comes from the smallness of $\gamma_i$ for large $i$. More specifically, the $1/\gamma_i$ scaling becomes ill-conditioned for large $i$, as it could be distorted by noise in the wave profile[11,12,21].

**Fundamental performance bound of the Slepian-Pollak imaging**

The relationship between the magnitude of $\gamma_i$ and the Slepian-Pollak imaging performance can be quantified within a quantum-information framework. The quantum Fisher information – more rigorously, the reciprocal of quantum Cramer-Rao bound – quantifies the maximum attainable accuracy in the unbiased estimation of the parameters of interest, regardless of the specific measurement process[2,32,33]. We derive the quantum Cramer-Rao bound for the estimation of each $i$th Slepian-Pollak coefficient for a confined and arbitrary optical source profile in the case of optimal measurement of scattered coherent light and retrieve the relationship between resolution, FOV, and required number of photons (photon budget) to achieve a desired resolution. The image resolution ($\Delta$) for 1D imaging (2D generalization is straightforward) can be roughly quantified by the size of the FOV, $D$, divided by the number of Slepian-Pollak coefficients that can be reliably estimated, $M$[19,21],

$$\Delta \approx \frac{D}{M} \text{ such that } P_{\text{tot}}|\gamma_M|^2 \sim 1 \qquad (1)$$

where $P_{\text{tot}}$ is the total number of scattered photons (see Supplementary Text S2 for the full derivation). The graphical representation of Eq. (1) is given in Fig. 2a where $\gamma_i$ are linearly interpolated.

It is immediately evident that, for large $D$, an impractically high photon budget is needed to increase the optical resolution beyond the conventional limit. Furthermore, it can be deduced that that the number of photons needed to achieve a desired resolution decreases proportionally to $D$, and that high resolution is possible, by increasing the number of involved photons, for a given $D$, and, conversely, by decreasing $D$, for a given photon budget. This is a consequence of the fact that the fast decay of the estimation accuracy for higher-order Slepian-Pollak coefficients can be significantly mitigated by the reduction of the FOV size, allowing to estimate coefficients for rapidly oscillating Slepian-Pollak functions with the same level of accuracy as those for slowly oscillating ones, by using an experimentally practical photon budget (see Supplementary Fig. S1).

**Implementation of the Slepian-Pollak imaging process**

To experimentally measure the complex amplitudes of the Slepian-Pollak coefficients of coherent light scattered from a confined object, we implement a shot-noise limited, spatial $N$-mode tomography[34,35], whose essential components are shown in Fig. 2b (see Methods section 1–2, Supplementary Figs. S2–4 for full description). A reconfigurable mask for the sequential tomography is placed at the Fourier plane to transform a sum of the reference and a target Slepian-Pollak component into a propagating plane wave, which is then focused on the detector by a lens. This measurement scheme can be compared with single-pixel homodyne detection but without the need for a separate reference light, because one of the strong Slepian-Pollak components is utilised as such.

The number of measured photons can be scaled by repeating the measurement specified by an $N$-by-$N$ diagonal matrix $R$, where the $k$th diagonal element, $R_{kk}$, represents the number of repetitions of the $k$th tomography ($k=1,…,N$) (Fig. 2c). Because of optical distortions present in realistic optical setups, the realisable imaging resolution can significantly differ from the ideal case. The experimentally obtainable resolution depends on the number of Slepian-Pollak coefficients that satisfy the relation:

$$P_{\sin} \frac{[\{\mathrm{Re}(T^H T)\}^{-1}]_{rr}}{[\{\mathrm{Re}(T^H R T)\}^{-1}]_{ii}} \geq 1 \qquad (2)$$

i.e. those that can be reliably estimated, where $P_{\sin}$, $T$, $H$, and $r$ are the measured photon number per a single tomography measurement without repetition, the $N$-by-$N$ transfer matrix of the optical system, its Hermitian transpose, and the index for the reference mode, respectively (see Supplementary Text S3). The condition in Eq. (2) can be satisfied by reducing the value of the denominator: one can obtain the target resolution by sufficiently repeating the shot-noise limited measurements (i.e. increasing the values of the elements of $R$) for a given transfer matrix, $T$, which represents the propagation and crosstalk of the optical setup. The transfer matrix is constructed by measuring the Slepian-Pollak coefficients of a single known object (a small dot for 2D, a thin line for 1D imaging). The actual Slepian-Pollak coefficients are estimated using a vector linear least mean square filter matrix[36], which is an inverse of the measured transfer matrix (see Methods section 3, Supplementary Fig. S5).

**Experimental demonstration of Slepian-Pollak imaging**

As proof-of-principle demonstrations, we perform Slepian-Pollak imaging of various nanoparticles of size <$0.8\lambda$ ($\lambda$ = 638 nm), which are fabricated on one face of sapphire cubes using focused electron beam induced deposition of platinum (see Methods section 4). The microscope part of Fig. 2b shows the total internal reflection (TIR) scattering configuration using an adjacent face of a sapphire cube, which effectively realise a radiation within the FOV (see Methods section 1, Supplementary Fig. S2). Four sets of Slepian-Pollak coefficients, for different illumination directions of the same object, are measured using various adjacent faces of the cubes. We measure and estimate 13 Slepian-Pollak coefficients for each object and for each illumination direction. A final image of an object is reconstructed as a squared magnitude of the coherent superposition of (TIR phase-compensated) Slepian-Pollak series with various illumination directions (see Supplementary Figs. S6 and S7 for the estimated Slepian-Pollak series for each illumination direction and raw data, respectively).

Figure 3 shows the experimentally obtained images of various nanoparticles. The imaging performance is good for objects of various shapes, in the absence of 4-fold rotational symmetry (compare #2 and #3), in the absence of mirror symmetry (compare #2 and #4), in the absence of a square grid (compare #1–4 and #5–8), and presence of curved features (#7–8). The images from experimental Slepian-Pollak imaging match the simulations well. In contrast, the calculated ideal conventional Fourier imaging with a numerical aperture (NA) of 0.9 (i.e. NA of the utilized objective lens) cannot resolve any of them (see Methods section 5).

We determine the imaging resolution based on a number of quantitative analyses. The spectra of the experimental images in Fig. 3 (last row) confirm that a significant portion is well extrapolated outside of the limited band (red circle) imposed by the NA of the objective lens. The yellow circles indicate the effective spectral band for the experimental Slepian-Pollak imaging, where the radius can be considered as representing an effective NA (see Methods section 5). We have achieved effective NA in the range from 2.43 to 3.6, 3.16 on average. Supplementary Fig. S8 shows the experimental demonstration of 1D Slepian-Pollak imaging, with an effective NA of about 3.53. The imaging performance can be also quantified with the overall Point Spread Function (PSF) of the instrument – from measurement to image reconstruction (Supplementary Text S4) [12,22]. Fig. 4a shows a PSF of our imaging system, compared to the PSF of the ideal Fourier imaging with NA = 0.9. The averaged full-width half-maximum of the PSFs over various positions, within the FOV, are ~$\lambda$/6 and ~$\lambda$/8, for 2D and 1D case, respectively (see Supplementary Fig. S9). Finally, by using sample #5 as a nanoscale Siemens star, we obtain a resolution of ~$\lambda$/7, as shown in Figure 4b, where we plot the intensity profile and its best-fit sine curve along a circumference of $\lambda$/5 diameter, the minimum diameter that fulfils the Rayleigh criterion between two spokes (20% saddle depth, or equivalently 10% modulation) [37]. Based on these three diverse criteria, we can claim an imaging resolution of around $\lambda$/7 for 2D imaging and $\lambda$/8 for 1D imaging.

In addition to image resolution, the Slepian-Pollak imaging can be quantified by the accuracy of each Slepian-Pollak coefficient estimation. Figure 4c compares the simulated and experimentally retrieved Slepian-Pollak coefficients – real and imaginary parts – for 2D imaging on all channels. It can be confirmed that the estimation accuracy is sufficiently good even for the weakest channel, where the normalized energy transfer ratio (i.e. $|\gamma_i/\gamma_1|^2$) is less than $10^{-5}$ and the crosstalk is stronger than the true signal (see Supplementary Fig. S5 for the filter). The channel-by-channel comparisons are shown in Supplementary Fig. S10 for both 2D and 1D imaging, which confirms the successful information transfer through a 1D Slepian-Pollak channel with an energy transfer ratio even lower than $10^{-6}$.

## Discussion

In this work, we reported the experimental demonstration of Slepian-Pollak imaging – a label-free far-field super-resolution imaging technique – with a resolution of $\lambda$/7 (2D imaging) and $\lambda$/8 (1D imaging) with the sole prior knowledge of the object size limit (<0.8$\lambda$), an accomplishment that all theoretical investigations deemed hardly feasible because of the extremely high measurement accuracy required. Our successful experimental implementation of Slepian-Pollak imaging, leveraged the ability to reliably measure both low and high order Slepian-Pollak coefficients of the source profile. We achieved this by first quantitatively characterising the experimental requirements based on a quantum-information framework, and then implementing a shot-noise-limited sequential single-pixel homodyne spatial mode tomography to measure the complex amplitudes of the Slepian-Pollak coefficients. The single-pixel homodyne tomography we used presents a clear benefit for the characterisation of the transfer matrix of the optical setup (or, equivalently, the vector linear filter): it allows a single measurement of a known object using many pixels simultaneously. Obtaining data from $L$ pixels simultaneously is equivalent to performing $L$ parallel tomographies with shifted objects, where the camera sensor is placed at the conjugate plane of the object (see Methods section 3).

We evaluate the performance of the Slepian-Pollak imaging also through the mutual information for each Slepian-Pollak channel (i.e. coefficients), showing a significant information transfer even through very weak channels (energy transfer ratio of $10^{-5}$–$10^{-6}$) in the presence of crosstalk. The choice of 13 and 6 Slepian-Pollak coefficients (measured and estimated) for 2D and 1D imaging, respectively, for each object and illumination direction, was dictated by practical considerations and allowed us to achieve deep experimental super-resolution imaging. While it is possible, in principle, to increase the number of coefficients by increasing the measurement repetitions, it becomes practically more challenging to achieve shot noise limited measurements. In Supplementary Fig. S11, we show how the filter performance in the case of 17 coefficients for 2D imaging becomes unsatisfactory even though we reach the required numbers of measurements repetition. This could be due to the non-zero scattering from surface roughness, especially from outside of the FOV, a known problem in particle localization[38] and mass photometry[39], which could be mitigated by measuring the surface roughness before preparing the sample[38,39].

It shall be noted that, while it may look complex, the realization of a Slepian-Pollak imaging setup simply requires the extension of a conventional microscope with a 4f optical system and a variable mask. Once assembled, the operation of a Slepian-Pollak microscope becomes straightforward and its calibration and transfer matrix quantification can be done using a standard object.

In conclusion, our experimental demonstration of Slepian-Pollak imaging, underpinned by the determination of the required number of measurement repetitions to retrieve each coefficient of the series and the development of a suitable vector linear filter, represents a milestone in label-free far-field super-resolution imaging. Our method enlarges the applicability of super-resolution label-free far-field imaging since the restriction of size limit can be considered as a widely acceptable requirement in nanotechnology and biomedical applications comparably to the assumptions utilized in previous experimental demonstrations[40]. Furthermore, the demonstrated method can be utilized in various imaging, metrology, spectroscopy, and LIDAR applications requiring accurate quantification of the spatial profile of light.

# Methods

## 1. Illumination on imaging objects, measurement of scattered light, estimation of Slepian-Pollak series, and reconstruction of an image

Supplementary Figs. S2a and S2b illustrate the total internal reflection illumination with a sapphire cube. A sapphire cube is a single crystal sapphire having an edge length of 7.5 mm, and two of the planes are c-normal, where nanoparticles are fabricated on one of the c-planes. Coherent visible light from a frequency stabilised diode laser (HÜBNER Photonics, Cobolt 08-01, $\lambda$=638 nm, linewidth < 0.1 pm) is utilised for the total internal reflection illumination on imaging objects. A free-space s-polarized laser beam is incident on the face of the sapphire cube, which is adjacent to the plane where nanoparticles are placed, with a nominal incidence angle of 45 degrees. This results in s-polarized beam illumination from the sapphire cube ($n_o$=1.765) on the imaging objects with an incidence angle of around 66 degrees, configuring total internal reflection illumination. The laser beam is weakly focused to position the beam waist roughly at the object plane. The monochromatic aberrations are roughly corrected with an additional cylindrical lens with appropriate curvature considering the beam path along the free-space and the sapphire cube. Incident power at the imaging object is roughly 20 mW, and the elliptical beam spot size is roughly 5 μm (short axis) and 12 μm (long axis).

The optical setup for the measurement of Slepian-Pollak coefficients is shown in Supplementary Fig. S2c. The sample (sapphire cube with imaging object), objective lens, and tube lens are placed on a piezo stage (Physik Instrumente, P-545.3R8S), which is fixed at a commercial inverted microscope platform (Nikon, Eclipse Ti). The piezo stage is utilised for the pre-positioning of the sample and the drift correction during the spatial mode tomography (see Methods section 2). The scattered light is collected by an objective lens with the numerical aperture of 0.9. The secondary laser is utilised to measure the longitudinal position of the sample during the pre-positioning process and is turned off during the spatial mode tomography (see Methods section 2). For the spatial mode tomography, a digital micromirror device (DMD, Texas instrument, DLP500YX) is placed roughly at the conjugate plane of the back focal plane of the objective lens (single-lens 2f system after the image plane of the object at the microscope side port). The pupil diameter of the objective lens at the DMD plane is around 4.4 mm. Complex amplitude masks of DMD are realised by utilising appropriate diffraction order where one super-pixel of the complex amplitude masks consist of four-by-four DMD pixels[41]. The DMD is aligned to realise Littrow configuration to minimise the distortion of spatial mode during the light propagation, such as the formation of an elliptic beam waist[42]. The inherent wavefront error of DMD (due to its imperfect flatness) is corrected as much as possible by characterising the error[43] and applying the phase-conjugated error to all masks used in the experiment. Eight images with different DMD masks are captured using a cooled SCMOS camera (TUCSEN, Dhyana 400BSI V2) to measure a Slepian-Pollak coefficient. Before the measurement of each Slepian-Pollak coefficient, the reference Slepian-Pollak component (strongest component among three components associated with the three largest $\gamma_i$) is chosen, and $I_a$, $I_b$, $\phi_{ba}$ are measured, which are the intensity at a pixel (digital number normalised by the pixel area, exposure time per frame, and the number of integrated frames) with the partial (a) DMD mask for the reference mode, the intensity at a pixel with the remainder (b) DMD mask for the reference mode, and phase difference between the optical fields at a pixel correspond to $I_b$ and $I_a$, respectively. To determine the $k$th (surely distorted) Slepian-Pollak coefficient, $c_k$, the intensities at a pixel of eight images ($I_{a,v}^{(k)}$ and $I_{b,v}^{(k)}$ where $v = 1, 2, 3, 4$) are measured, and the following equation is used,

$$c_k = \sum_{v=1}^{4} \frac{1}{4} \left[ \frac{(-i)^v I_{a,v}^{(k)}}{\sqrt{I_a}} + e^{i\phi_{ba}} \frac{(-i)^v I_{b,v}^{(k)}}{\sqrt{I_b}} \right]. \tag{3}$$

The DMD masks for measuring $I_a$, $I_b$, $\phi_{ba}$, $I_{a,v}^{(k)}$, and $I_{b,v}^{(k)}$ ($v = 1, 2, 3, 4$) are illustrated in Supplementary Fig. S3. We confirmed that the photon number from the reference Slepian-Pollak component per frame is sufficiently large near the nominal centre pixel of the camera for all cases (i.e. the shot noise of the reference Slepian-Pollak component is much larger than the camera readout noise and dark current), which is a necessary condition to be a shot noise limited measurement. Then, shot noise limited performance is maintained for the entire repetitive measurement by drift correction during the measurement (see Methods section 2). The nominal centre pixel is where the optical axis of the optical system crosses the camera sensor, which also corresponds to the rough centre of the direct image of the object (see Methods section 2). For all measurements, the exposure time per frame (i.e. each repetition) is chosen to avoid the saturation of pixel full well depth, and the sufficient number of frames are integrated, which is determined by the measured transfer matrix, $T$ (see Methods section 3).

The set of Slepian-Pollak coefficients for an imaging object under a specific illumination at the object plane is estimated as $\mathbf{c}_e = W\mathbf{c}_m$ where $\mathbf{c}_e$, $W$, and $\mathbf{c}_m$ are the $N$-by-1 column vector of estimated Slepian-Pollak coefficients at the object plane, the vector linear filter, and the $N$-by-1 column vector of measured Slepian-Pollak coefficients, respectively, and $N$ is the number of measured and estimated coefficients (see Methods section 3). Only one $\mathbf{c}_m$ (thus only one $\mathbf{c}_e$) is chosen for further imaging process by choosing the correct centre pixel for the Slepian-Pollak series expansion. The correct centre pixel is determined after the tomographic measurement among the nominal centre pixel and its peripheral pixels such that the energy of the estimated Slepian-Pollak series (at the object plane) is the smallest at the correct centre pixel.

The image of an object (also can be considered as a map of square magnitude of polarizability) is reconstructed as the following equation,

$$\text{Image} = \left| \sum_{l=1}^{4 \text{ (or 2)}} \left[ e^{-i(\mathbf{k}_{\text{TIR},l} \cdot \mathbf{r} + \phi_l)} \sum_{j=1}^{N} c_{e,j} S_j(\mathbf{r}) \right] \right|^2 \tag{4}$$

where $\mathbf{k}_{\text{TIR},l}$ is the wavevector of $l$th illumination light and $\phi_l = \angle \left[ \int_{|\mathbf{r}| \leq R_O} \left( \sum_{k=1}^{N} c_{e,l}(k) S_k(\mathbf{r}) \right) d\mathbf{r} \right]$ is a $l$th global phase correction factor. In other words, the image of an object is reconstructed by the square magnitude of coherent summation of four (for two-dimensional imaging) or two (for one-dimensional imaging) sets of Slepian-Pollak coefficients associated with various illumination directions where TIR illumination phases are compensated. Various illumination directions are realised by utilising four or two adjacent faces of the sapphire cube compared to the face where the imaging objects are placed.

## 2. Pre-positioning of an imaging object before the tomographic measurement and drift correction during the tomographic measurement

Before the tomographic measurement of the Slepian-Pollak series, we measure the three-dimensional position of the object and move its centre to the nominal origin (i.e. $(x, y, z) = (0, 0, 0)$) with the piezo stage. First, the (longitudinal) $z$-coordinate of the centre position is assumed to be the same as the surface of the sapphire cube. It is determined by maximum likelihood estimation (MLE) using the modified direct image (please refer to the last paragraph of this section for more detail about the modified image) of the beam spot reflected from the surface of the sapphire cube illuminated from the secondary laser (see Supplementary Fig. S2 for the setup configuration and spatial coordinate). The probability distributions (i.e. modified images) as a function of $z$-coordinate are pre-recorded for the MLE. Then, the non-zero $z$-coordinate is compensated to $z=0$ using the piezo stage. Next, the object's centre's $x$- and $y$-coordinates are determined by the first moment of the intensity-truncated direct image (specifically, min[$I(x,y)$, $0.1\max\{I(x,y)\}$] where $I(x,y)$ is the direct image). Non-zero $x$- and $y$-coordinates are compensated to $x=0$ and $y=0$ using the piezo stage where the sample is placed. This realises the direct image of the object forms near the nominal centre pixel of the camera. The secondary laser is turned off after the pre-positioning.

During the tomographic measurement, mechanical drifts of the optical components are compensated in real time to realise shot noise limited (not limited by the drift) measurement. The drift compensation and the tomographic measurement of Slepian-Pollak components are performed alternatively (switches every 2-3 seconds) until the end of the tomographic measurement. Although drifts from various optical components lead to different effects on the scattering pattern, they can be well approximated to an equivalent displacement of the sample for small drifts. The $x$-, $y$-, and $z$-directional displacements are determined by MLE using modified direct images (please refer to the last paragraph of this section for more detail about the modified direct image) of the imaging object. The probability distributions (i.e. modified image) for various $z$-displacements (without $x$- and $y$-displacements) of the object are pre-recorded, and the probability distributions for various $x$- and $y$-displacement for every $z$-displacement are calculated right after the aforementioned pre-positioning before the drift compensation for the MLE. Non-zero $x$-, $y$-, and z-components of displacements are compensated using the piezo stage where the sample is placed.

For a good enough estimation of three-dimensional position and displacement, we optically modify the direct image of the secondary laser beam spot and imaging object by utilising a purposely designed DMD mask such as in Ref. 44 because the focused image of a point-like object is not suitable for the $z$-directional position or displacement estimation[33]. Supplementary Fig. S4a shows the DMD mask for this purpose: spatial multiplexing of two masks – one changes orbital angular momentum amount of +2 order (outer part) while another (inner part) does not. The calculation shows that the resulting modified image of a point-like object (shown in Supplementary Fig. S4b) provides per-photon Cramer-Rao bound of $x$-, $y$-, and $z$- displacements estimation of $0.1\lambda$, $0.14\lambda$, and $0.32\lambda$ (in terms of standard deviation), respectively, which is reasonably good performance compared to Ref. 33, and sufficient for our purpose. The good enough Cramer-Rao bound for the longitudinal position estimation can be understood by the difference of the Gouy phase dynamics along the longitudinal axis between beams with different orbital angular momentum[45]. Supplementary Fig. S4c shows the object's position during the tomographic measurement, which shows the object's position is well maintained. Supplementary Fig. S4d shows the accumulated counter-movement of the piezo stage (inferred by the voltages applied to piezo actuators) during the tomographic measurement of Slepian-Pollak coefficients, which is around a few hundred nanometres. The piezo stage is operated in the open-loop mode to avoid its own mechanical vibration occurring in the closed-loop mode.

## 3. Characterisation of the vector linear filter, the transfer matrix, and the required number of repetitions for the measurement

We consider a sequential spatial $N$-mode tomography where $N$ is the number of Slepian-Pollak coefficients to be measured and estimated. For a realistic optical system, an $N$-by-1 measured set of coefficients of some basis (distorted Slepian-Pollak basis), $\mathbf{c}_m$, can be represented as $\mathbf{c}_m = T\mathbf{c}_o$ where $T$ is the $N$-by-$N$ transfer matrix and $\mathbf{c}_o$ is the $N$-by-1 vector of actual Slepian-Pollak coefficients of the source. In this case, an $N$-by-1 vector linear minimum mean square error estimator, $\mathbf{c}_e = W\mathbf{c}_m$, can be considered as a solution for Slepian-Pollak imaging where $W = (T^{-1})_{\text{lms}}$ is the $N$-by-$N$ least mean squares (LMS) filter matrix. The LMS filter matrix can be characterised by

$$W = \Omega_{\text{om}} \Omega_{\text{mm}}^{-1} \qquad (5)$$

where $\Omega_{\text{om}}$ is the $N$-by-$N$ cross-correlation matrix between various $\mathbf{c}_o$ and various $\mathbf{c}_m$, and $\Omega_{\text{mm}}$ is the $N$-by-$N$ autocorrelation matrix of various $\mathbf{c}_m$[36].

We measure $\mathbf{c}_m$ for known $\mathbf{c}_o$ for the characterisation of the LMS filter. Let $N$-by-$L$ matrix $C_m$ is the collection of measured vectors (i.e. $C_m = [\mathbf{c}_m^{(1)}, \mathbf{c}_m^{(2)}, ..., \mathbf{c}_m^{(L)}]$) and $C_o$ is the corresponding $N$-by-$L$ matrix, a collection of known Slepian-Pollak series of objects under an illumination (i.e. $C_o = [\mathbf{c}_o^{(1)}, \mathbf{c}_o^{(2)}, ..., \mathbf{c}_o^{(L)}]$). Then, $\Omega_{\text{om}} = C_o C_m^H$ and $\Omega_{\text{mm}} = C_m C_m^H$, where $H$ in the superscript denotes the Hermitian transpose. Because an *N*-by-1 column vector can be determined from a single pixel on the camera for a tomographic measurement, we could construct $C_m$ by obtaining data from *L* number of (around 100) pixels near the nominal centre pixel of the camera from a single set of tomography. We measure only one known calibration sample for each type of imaging: a dot with an 80 nm diameter for two-dimensional (2D) imaging and a line with an 80 nm width for one-dimensional (1D) imaging. Then, $C_o$ effectively becomes a collection of $\mathbf{c}_o$ corresponds to dots placed at various positions in the object plane.

The accuracy of the LMS filter matrix from Eq. (5) has yet to be determined because $\mathbf{c}_m$ always includes noise. Fortunately, the noise is controllable by adjusting the number of repetitions of measurements (i.e. integrating frames) for a shot noise-limited case. Now, we determine the required number of repetitions. A measured vector can be represented as $\mathbf{c}_m = T\mathbf{c}_o + \mathbf{c}_n$ where $\mathbf{c}_n$ is the *N*-by-1 noise vector. Then, $\Omega_{om} = \Omega_{oo}T^H$ and $\Omega_{mm} = T\Omega_{oo}T^H + \Omega_{nn}$, where $\Omega_{oo}$ and $\Omega_{nn}$ are the *N*-by-*N* autocorrelation matrix of various $\mathbf{c}_o$ and the *N*-by-*N* autocorrelation matrix of $\mathbf{c}_n$, respectively. By assuming $\|T\Omega_{oo}T^H\|_F \gg \|\Omega_{nn}\|_F$ where $\|\bullet\|_F$ is the Frobenius norm of $\bullet$, it can be approximated that $\Omega_{mm}^{-1} \approx (T^{-1})^H \Omega_{oo}^{-1} T^{-1} + (T^{-1})^H \Omega_{oo}^{-1} T^{-1} \Omega_{nn} (T^{-1})^H \Omega_{oo}^{-1} T^{-1}$. Then, from Eq. (5) and using $T^{-1} \approx W_0$ where $W_0$ is the noiseless LMS filter, the LMS filter can be approximated as

$$W \approx W_0 \left( I + \Omega_{nn} W_0^H \Omega_{oo}^{-1} W_0 \right) \qquad (6)$$

where *I* is the *N*-by-*N* identity matrix. The autocorrelation matrix of $\mathbf{c}_n$ can be represented as $\Omega_{nn} = \sigma_n^2 R^{-1}$ where $\sigma_n^2$ is the noise variance for measurement without repetition, and *R* is the diagonal matrix whose *k*th diagonal component, $R_{kk}$, is the number of repetitions for *k*th mode tomography. We always measure around 1500 photons per frame per pixel, which is mainly from the reference mode (i.e. *k*=1 for a small dot and thin line sample). Then, it can be approximated that $\Omega_{mm,11} = |T_{11}|^2 \Omega_{oo,11} = 1500$ by assuming $|T_{11}| \gg |T_{1p}|$ (*p* = 2,…*N*) and considering $\sigma_n^2 = 2$ (see Supplementary Text S3). We determined the required number of repetitions such that

$$\begin{cases} \partial \|W\|_{\mathrm{F}} / \partial R_{kk} = \text{constant for all } k \\ \qquad\qquad \text{and} \\ \left\| \Omega_{\mathrm{nn}} W_0^H \Omega_{\mathrm{oo}}^{-1} W_0 \right\|_{\mathrm{F}} = \varepsilon \|I\|_{\mathrm{F}} \end{cases}. \qquad (7)$$

where error factor $\varepsilon$ should be kept small (e.g. $\varepsilon=0.1$). Using an iterative scheme, we first assume $W_0 = \Omega_{\mathrm{om}} \Omega_{\mathrm{mm}}^{-1}$ (from Eq. (5)) with an initial guess of $R$. However, we note that the LMS filter and required $R$ can be obtained accurately at the first iteration if one chooses sufficiently large values of the initial guess of $R_{kk}$, for all $k$. In addition to the noise, singular values of $\Omega_{\mathrm{oo}}$ should be large enough so that $(\Omega_{\mathrm{oo}})^{-1}$ can be well defined. It can be confirmed that singular values of $C_\mathrm{o}$ (hence, singular values of $\Omega_{\mathrm{oo}}$ as well) are reasonably well distributed in our case (i.e. collection of shifted dots or lines). Supplementary Fig. S5a shows the characterised $R_{kk}$ following Eq. (7) and the utilised $R_{kk}$ for 2D imaging experiments for all $k$. Supplementary Fig. S5b show the experimentally characterised LMS filter matrix following this procedure.

Supplementary Fig. S5c shows the average residue error for $i$th Slepian-Pollak components (i.e. $\left[(C_\mathrm{o} - WC_\mathrm{m})(C_\mathrm{o} - WC_\mathrm{m})^H\right]_{pp} / \left[C_\mathrm{o} C_\mathrm{o}^H\right]_{pp}$) for all $i$, which are all small. Finally, we calculate achievable accuracy for the estimation of Slepian-Pollak coefficients of unknown objects using the experimentally characterised LMS filter and the utilised $R$. Supplementary Fig. S4d shows the expected signal-to-noise ratio (i.e. actual signal variance divided by noise variance) of $i$th Slepian-Pollak coefficient estimation with a repetition specified by $R$, $\mathrm{SNR}_{R,i}$, using Eq. (S28) in Supplementary Information and assuming $T=W^{-1}$. Because $\mathrm{SNR}_{R,i}$ is good enough for all $i$, it can be concluded that the utilised $R$ is not only sufficient for the accurate characterisation of the filter but also sufficient for the estimation of $i$th Slepian-Pollak coefficient of unknown source under observation for all $i$, for the utilised optical setup.

## 4. Preparation of imaging objects

A fine-polished crystalline sapphire cube (edge length: 7.5 mm) is utilised as substrate for the imaging nanoparticles and prism for the total internal reflection illumination simultaneously. Before the fabrication of nanoparticles, the sapphire cube is cleaned using SC1 (standard clean

1) process with ammonium hydroxide, hydrogen peroxide, and deionised water. We deposit a smooth conductive layer (2-nm thick NbTiN) for good enough electric conductivity for the following nanoparticle fabrication. We fabricate platinum nanoparticles of various shapes on a *c*-plane of fine-polished sapphire cube using focused electron beam induced deposition (FEBID) of platinum (precursor gas: $C_5H_4CH_3Pt(CH_3)_3$). The nominal thickness of the fabricated nanoparticles is 32 nm. The NbTiN layer is oxidised after FEBID for optical transparency using brief reactive ion etching with oxygen.

## 5. Simulation and calculation for Slepian-Pollak imaging and ideal Fourier imaging, and quantification of effective numerical aperture.

The simulation result for the Slepian-Pollak imaging shown in Fig. 3 in the main text is obtained using finite-difference time-domain (FDTD) simulation followed by Fourier optic calculation. Total internal reflection illumination is realised in the FDTD simulation model with a total-field scattered-field source with the designed angle of incidence (i.e. around 66 degrees) illuminated from the sapphire substrate on which the platinum nanostructure is placed. The scattered electric and magnetic near-field profiles are collected for a large enough area in the free-space region near the imaging object on the sapphire cube. The near-field data is transformed to far-field in the angular spectrum, and it is further mapped to confined rectilinear coordinates, which are assumed to be the back focal plane of the objective lens with a numerical aperture of 0.9. As in the experiment, only horizontally polarised components are utilised. The rest of the optical path is modelled in the Fourier optic calculation using similar parameters compared to the experiment, such as optical path difference introduced by lens and propagation distances. To obtain the *i*th Slepian-Pollak coefficient, a complex conjugation of $S_i$ is used as a hypothetical DMD mask (without mixing with the reference mask, which is different from the experiment, as shown in Supplementary Fig. S2). Similar to the experiment, the amplitude value at the centre of the camera plane is utilised to determine each coefficient, and an *N*-by-1 Slepian-Pollak coefficient, $c_m$, is retrieved for each nanoparticle and for each illumination direction. The rest of the image formation is same as the experimental procedure shown in Methods section 1.

The images from ideal conventional imaging shown in Fig. 3 in the main text are low-pass filtered binarised scanning electron microscopy (SEM) images of nanoparticles where the threshold spatial frequency is $0.9k_0$ (= $0.9(2\pi)\lambda^{-1}$).

To quantify the effective numerical aperture, $NA_{eff}$, for each Slepian-Pollak image, first, the fidelity (between 0 to 1) between the binarised SEM image and the experimental Slepian-Pollak image is calculated. Then, find the effective threshold spatial frequency (i.e. $NA_{eff} \cdot k0$) for the low-pass filter for ideal conventional Fourier imaging, which gives the same fidelity between the binarised SEM image and the ideal conventional Fourier imaging with the $NA_{eff}$ compared to the fidelity for the Slepian-Pollak imaging case.

## Acknowledgements

The authors thank Shuyu Dong and Cesare Soci for the deposition of the smooth conductive layer (NbTiN thin film) for the imaging object fabrication and Jin-Kyu So, Anton Vetlugin, Yijie Shen, Nikitas Papasimakis, and Eng Aik Chan for useful discussions.

**Funding**

This work was supported by the Singapore National Research Foundation (Grant No. NRF-CRP23-2019-0006) and the Engineering and Physical Sciences Research Council UK (Grants No. EP/T02643X/1).

**Author contribution**

TC and NIZ conceived the idea, TC developed the methodology, assembled the apparatus, and conducted measurements. All authors contributed to the discussion and interpretation of the results. NIZ and GA supervised and administrated the project and cross-edited the original draft prepared by TC.

**Competing interests**

The authors declare no competing interests.

**Figures**

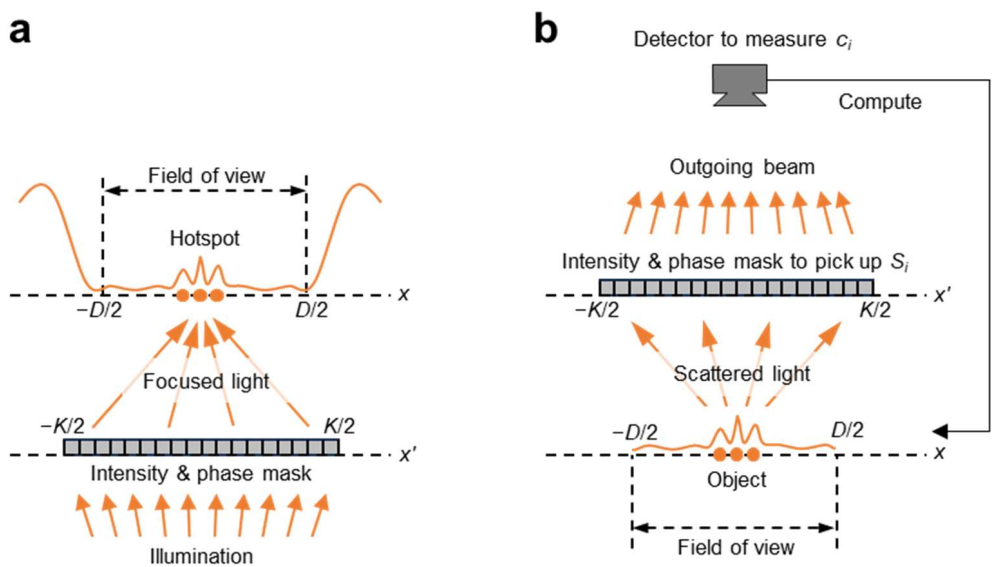

**Figure 1. Concept of Slepian-Pollak imaging as a reverse process of superoscillatory hotspot generation.** (a) Generation of a superoscillatory hotspot. A tailored source profile is created at the Fourier plane, within the limited band ($K$), by an intensity and phase mask and used to generate an arbitrary wave profile containing deeply subwavelength features, within the field of view ($D$) at the object plane. (b) Slepian-Pollak imaging. An arbitrary source profile within the field of view ($D$) at the object plane, originated from scattered light from an object and containing deeply subwavelength features, generates a unique wave profile in the Fourier plane, within the limited band ($K$). By measuring the Slepian-Pollak coefficients of the field amplitude at the Fourier plane, the source profile at the object plane can be estimated. The source-wave relationships for two cases form reciprocal configurations.

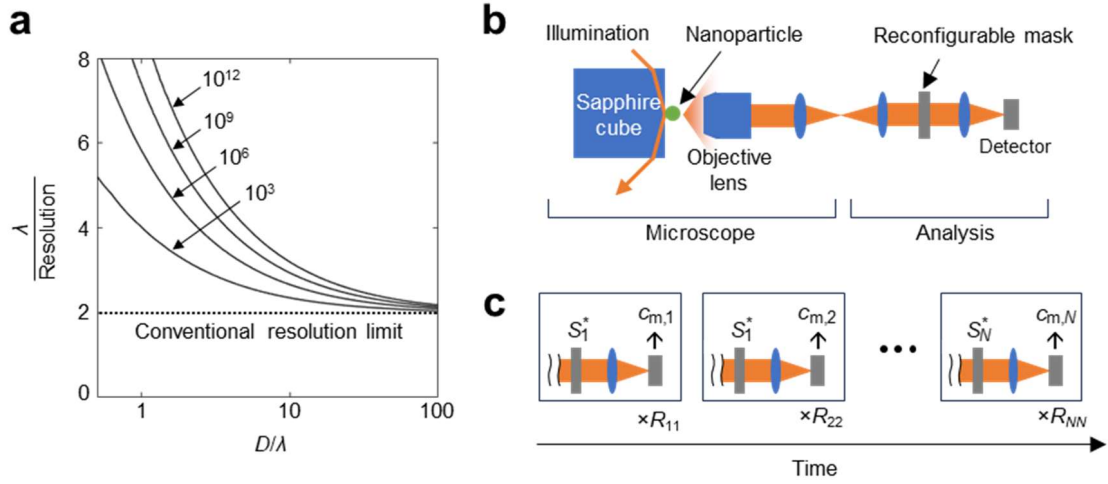

**Figure 2. Slepian-Pollak imaging limits and its experimental implementation.** (a) Fundamental bounds of Slepian-Pollak imaging performance. Achievable optimal resolution as a function of the size of the field of view (*D*) and total number of scattered photons (representative curves for $10^3$, $10^6$, $10^9$, and $10^{12}$ photons). (b) A simplified schematic of the experimental Slepian-Pollak imaging setup. Light illuminates an object in total internal reflection and the scattered light is collected by a microscope. The Slepian-Pollak coefficients at the Fourier plane are measured by spatial-mode tomography, with a reconfigurable mask. (c) Sequential tomography. A vector of Slepian-Pollak coefficients is obtained after *N*-mode tomography, where the *k*th tomography is repeated $R_{kk}$ times (*k*=1,…,*N*).

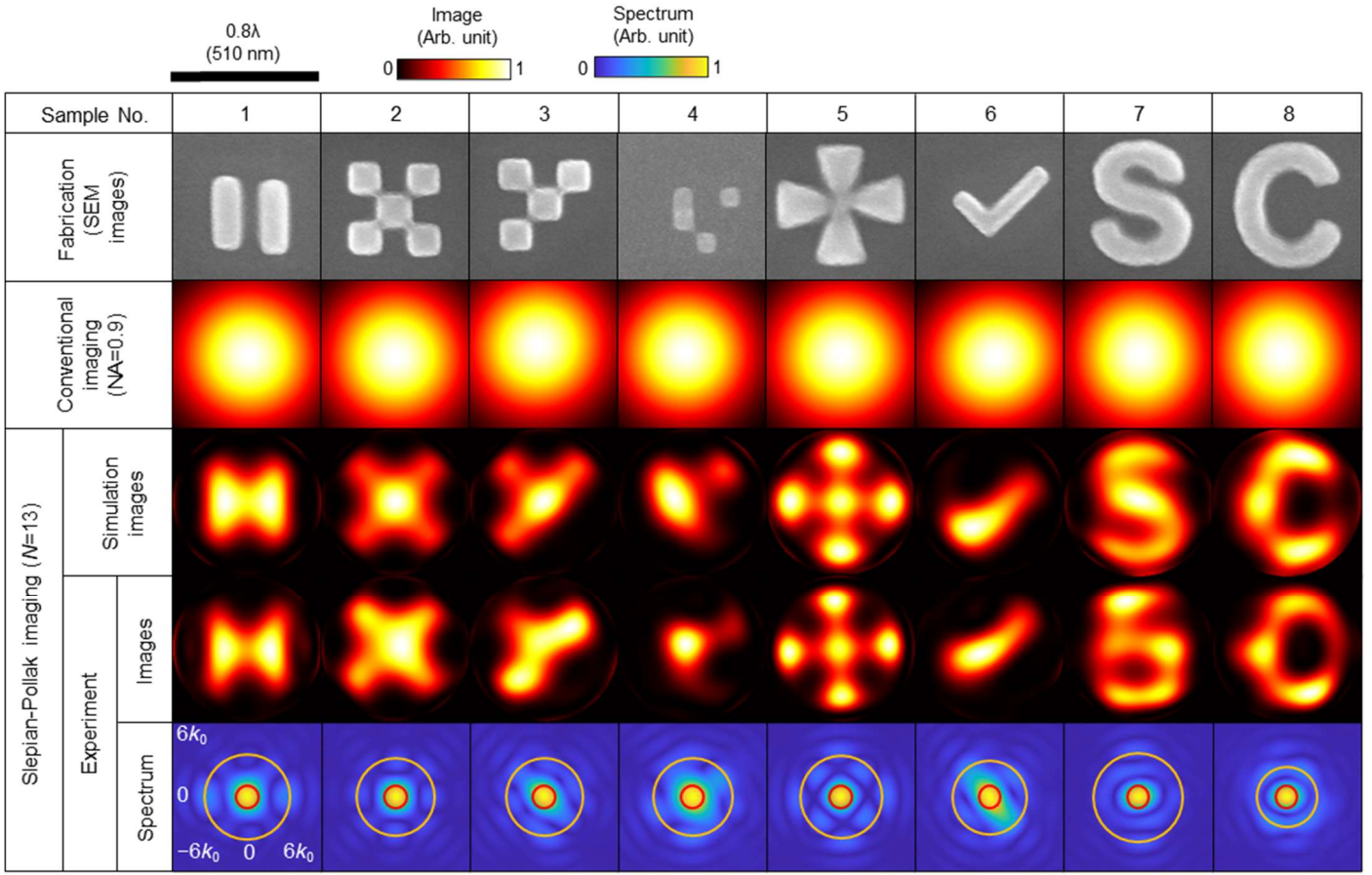

**Figure 3. Experimental demonstration of Slepian-Pollak imaging.** SEM images of nanoscale objects of various shapes, sizes and symmetries (first row from top) are compared with the calculated images for an ideal conventional Fourier imaging with numerical aperture NA=0.9 (second row), the simulated Slepian-Pollak images (third row) and the experimental Slepian-Pollak images (forth row), demonstrating the significant improvement in imaging performance brought about by the Slepian-Pollak imaging over the NA=0.9 Fourier imaging. The Spectra of the experimentally obtained images (fifth row) reveal an effective numerical aperture of 3.16.

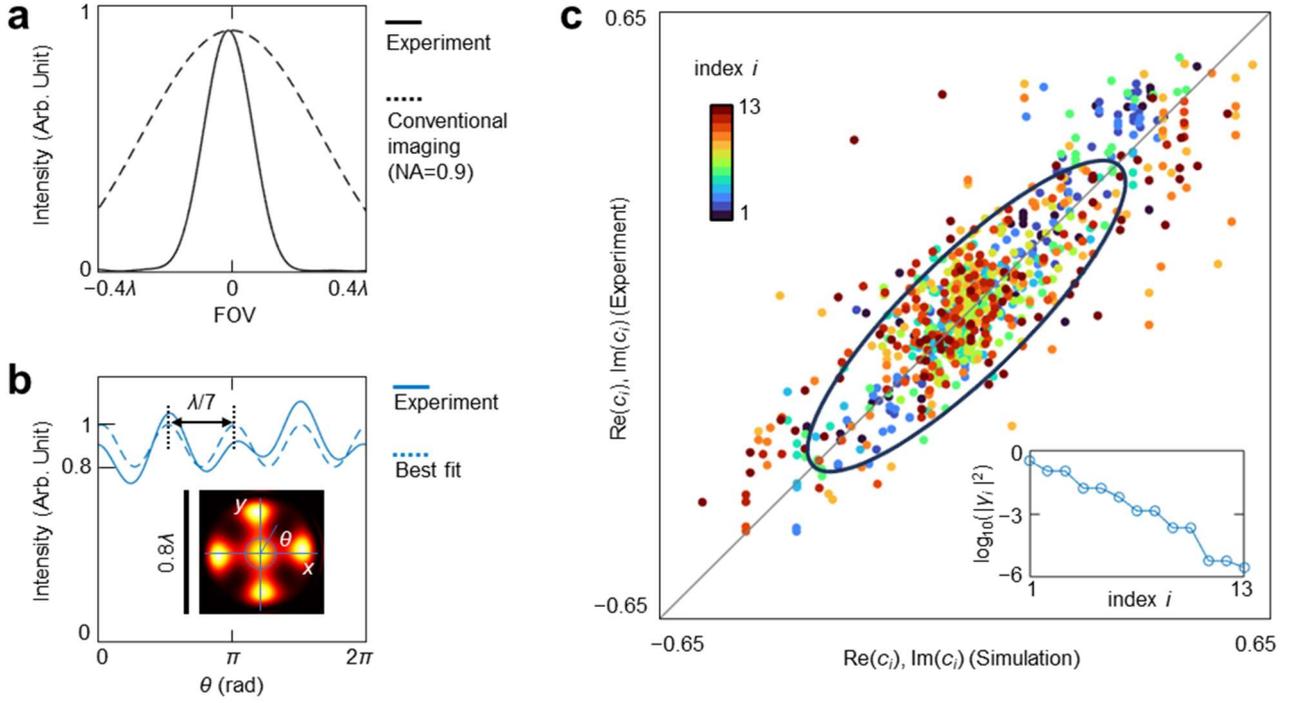

**Figure 4. Experimental resolution of the Slepian-Pollak imaging and coefficients retrieval accuracy.** (a) The representative point spread function of Slepian-Pollak imaging (full-width half-maximum ~$\lambda/6$) is compared to that of an ideal conventional Fourier imaging with numerical aperture NA=0.9. (b) Resolution test using a nanoscale Siemens star. The light intensity across the circumference with a minimum diameter that satisfies the Rayleigh criterion on average, reveals the ability to distinguish two spokes having a centre-to-centre distance of ~$\lambda/7$. (c) Real and imaginary parts of the experimentally measured Slepian-Pollak coefficients of Fig. 3 are compared with those obtained from simulations, showing a good estimation accuracy for both strong and weak channels. The coefficients from each channel are plotted with different colours. The continuous curve represents the $2\sigma$ ellipse of the covariance matrix. The value of $|\gamma_i|^2$ plotted in the inset quantifies the decrease of signal strength of Slepian-Pollak mode with the increase of the order $i$.

# Supplementary Information

## Supplementary Text

**S1. Generation of a superoscillatory hotspot and Slepian-Pollak imaging form reciprocal configurations.**

In figure 1 in the main text, we consider a wave propagation from one confined region to another confined region. Equivalently, one can consider an effective optical system with a confined region of multimode input and output ports (waves outside the confined region can be considered a loss). First, in Fig. 1b in the main text, output wave is located at the Fourier plane and the input source is located at the object plane (i.e. forward propagation). In this case, Slepian-Pollak functions represent normal modes of wave propagation because they are the eigenfunctions of finite Fourier transform[1,2]. This can be expressed as

$$\int_{|x|\leq D/2} G(x',x) S_i(x) dx = \gamma_i S_i(x') \quad \text{(for } |x'| \leq K/2\text{)} \tag{S1}$$

where $G$, $S_i$, and $\gamma_i$ are the Green's function, the $i$th Slepian-Pollak function ($i=1,2,3,\ldots$), and the complex amplitude transfer ratio of $i$th Slepian-Pollak function from one confined plane to another, respectively. The Slepian-Pollak functions, $S_i(x)$ and $S_i(x')$, satisfy the orthonormality condition,

$$\int_{|x|\leq D/2} S_i(x) S_j(x) dx = \delta_{ij}$$
$$\text{and} \tag{S2}$$
$$\int_{|x'|\leq K/2} S_i(x') S_j(x') dx' = \delta_{ij}$$

Multiplying $S_j(x')$ on both sides of Eq. (S1) followed by applying the orthonormality condition of Slepian-Pollak functions yields

$$\int_{|x'|\leq K/2}\int_{|x|\leq D/2} G(x',x)S_i(x)dx S_j(x')dx' = \gamma_i\delta_{ij}. \quad (S3)$$

Using the reciprocity of Green's function (i.e. $G(x',x)=G(x,x')$)[3], Eq. (S3) yields

$$\int_{|x|\leq D/2}\left[\int_{|x'|\leq K/2} G(x,x')S_j(x')dx'\right]S_i(x)dx = \int_{|x|\leq D/2}\left[\sum_k c_k S_k(x)\right]S_i(x)dx = \gamma_i\delta_{ij}, \quad (S4)$$

which yields $c_k=\gamma_i\delta_{ij}\delta_{ik}$ due to the orthonormality condition. Therefore,

$$\int_{|x'|\leq K/2} G(x,x')S_j(x')dx' = \gamma_j S_j(x) \quad \text{(for } |x|\leq D/2\text{)}. \quad (S5)$$

Equation (S5) implies that the Slepian-Pollak functions are also the normal modes for the backward wave propagation shown in Fig. 1a in the main text. We note that modal version of scattering amplitude, $f(S_i,S_j)$, satisfy the Helmholtz reciprocity (i.e. $f(S_i,S_j)=f(-S_j,-S_i)=\gamma_i\delta_{ij}$) (following the convention in Ref. 4) by observing Eq. (S1) and Eq. (S5). This is a direct consequence of the reciprocity in Green's function, as shown in the above formulation.

We define two optical situations as reciprocal configurations if the input source profile of one situation is the same as the output wave profile of another situation following the description in Ref. 5. In a superoscillatory hotspot generation (Fig. 1a in the main text), source profile of $\sum_i \gamma^{-1} c_i S_i(x')$ result in the wave profile (i.e. superoscillatory hotspot) of $\sum_i c_i S_i(x)$ from Eq. (S5). In Slepian-Pollak imaging (Fig. 1b in the main text), the source profile of $\sum_i c_i S_i(x)$ (i.e. same as the superoscillatory hotspot) results in the wave profile of $\sum_i \gamma_i c_i S_i(x')$ from Eq. (S1). Therefore, two situations form reciprocal configurations.

For an image reconstruction in Slepian-Pollak imaging, let us assume the measured wave profile is $\sum_i d_i S_i(x')$ without knowing the source profile. The imaging task is determining $c_i$ ($i=1,\ldots,N$) in a series $\sum_i c_i S_i(x)$ to reconstruct the source profile. This task can be expressed as solving a Fredholm integral equation,

$$\int_{|x|\leq D/2} G(x',x)\left[\sum_i c_i S_i(x)\right]dx = \sum_i d_i S_i(x') \quad \text{(for } |x'|\leq K/2\text{)}, \tag{S6}$$

to find the coefficient $c_i$. This task is straightforward from Eq. (S1), which yields $\sum_i \gamma_i c_i S_i(x') = \sum_i d_i S_i(x')$. By applying the orthonormality condition, it can be concluded that $c_i = d_i/\gamma_i$ for all $i$.

## S2. Derivation of the fundamental resolution limit of Slepian-Pollak imaging from the quantum Cramer-Rao bound for the estimation of Slepian-Pollak coefficients.

We first derive the $N$-by-$N$ quantum Fisher information matrix, $I(c_1,\ldots,c_N)$, and quantum Cramer-Rao bound for each parameter, $\sigma^2_{\text{QCRB},i}$ ($i=1,\ldots,N$), for the simultaneous estimation of $N$ number of Slepian-Pollak coefficients of the source profile by measuring Slepian-Pollak coefficients of its radiation following[6-8].

In general, the ($p$, $q$) component of the quantum Fisher information matrix for multi-parameter estimation is defined by

$$I_{pq} = \frac{1}{2}\text{Tr}\left[\rho\left(L_p L_q + L_q L_p\right)\right] \tag{S7}$$

where $L_r$ ($r = p, q$) and $\rho$ are the symmetric logarithmic derivative of $\rho$ with respect to the $r$th parameter, $\theta_r$, and the density operator representing the quantum state of the system, respectively[6]. $L_r$ is implicitly defined as

$$\rho L_r + L_r \rho = 2\partial_r \rho \tag{S8}$$

where $\partial_r[\bullet] = \partial[\bullet]/\partial\theta_r$ [6]. For the following derivation, we consider the wave radiated from the source of interest as a separable pure state representing an $N$-mode coherent state[8], which applies to our case. The density matrix becomes $\rho = |\psi\rangle\langle\psi|$ where $|\psi\rangle = |\alpha_1\rangle \otimes \cdots \otimes |\alpha_N\rangle$ and $|\alpha_k\rangle$ is the state vector for $k$th single mode. Using the condition for the pure state, the condition for unitary transformation (i.e. $\partial_r(\langle\psi|\psi\rangle) = 0$), and its consequences (i.e. $\text{Re}(\langle\psi|\partial_r\psi\rangle) = 0$ and $\langle\psi|L_r|\psi\rangle = 0$ )[7], the equation $(\rho L_p + L_p \rho)(\rho L_q + L_q \rho) + (\rho L_q + L_q \rho)(\rho L_p + L_p \rho) = 4\partial_p \rho \cdot \partial_q \rho + 4\partial_q \rho \cdot \partial_p \rho$ yields

$$I_{pq} = \frac{1}{2}\langle\psi|L_p L_q + L_q L_p|\psi\rangle = 4\text{Re}\left[\langle\partial_p\psi|\partial_q\psi\rangle\right] - 4\text{Re}\left[\langle\psi|\partial_p\psi\rangle\langle\partial_q\psi|\psi\rangle\right]. \quad (S9)$$

Using the condition for the separable state, Eq. (S9) becomes

$$I_{pq} = 4\text{Re}\left(\sum_{k=1}^{N}\langle\partial_p\alpha_k|\partial_q\alpha_k\rangle + \sum_{k,l=1(k\neq l)}^{N}\langle\partial_p\alpha_k|\alpha_k\rangle\langle\alpha_l|\partial_q\alpha_l\rangle\right) \\ - 4\text{Re}\left(\sum_{k,l=1}^{N}\langle\alpha_k|\partial_p\alpha_k\rangle\langle\partial_q\alpha_l|\alpha_l\rangle\right), \quad (S10)$$

which can be further reduced to

$$I_{pq} = 4\sum_{k=1}^{N}\left[\text{Re}(\langle\partial_p\alpha_k|\partial_q\alpha_k\rangle) - \text{Re}(\langle\alpha_k|\partial_p\alpha_k\rangle\langle\partial_q\alpha_k|\alpha_k\rangle)\right]. \quad (S11)$$

Using the Fock state basis representation of $|\alpha_k\rangle$, i.e. $|\alpha_k\rangle = e^{-|\alpha_k|^2/2}\sum_{n=0}^{\infty}\alpha_k^n(n!)^{-1/2}|n\rangle$ where $\alpha_k$ is the eigenvalue of the annihilation operator, which is the same as the complex amplitude of $k$th mode[8],

$$|\partial_r \alpha_k\rangle = -\text{Re}(\alpha_k \partial_r \alpha_k^*) e^{-|\alpha_k|^2/2} \sum_{n=0}^{\infty} \frac{\alpha_k^n}{(n!)^{1/2}} |n\rangle + \partial_r \alpha_k e^{-|\alpha_k|^2/2} \sum_{n=0}^{\infty} \frac{n\alpha_k^{n-1}}{(n!)^{1/2}} |n\rangle. \quad \text{(S12)}$$

Using $\sum_{n=0}^{\infty} |\alpha_k|^{2n} (n!)^{-1} = e^{|\alpha_k|^2}$, $\sum_{n=0}^{\infty} (n+1)|\alpha_k|^{2n} (n!)^{-1} = (1+|\alpha_k|^2) e^{|\alpha_k|^2}$, and the orthonormality of Fock states,

$$\langle \partial_p \alpha_k | \partial_q \alpha_k \rangle = \text{Re}(P)\text{Re}(Q) - \text{Re}(P)Q^* - \text{Re}(Q)P + \partial_p \alpha_k^* \partial_q \alpha_k (1+|\alpha_k|^2) \quad \text{(S13)}$$

where $P = \alpha_k \partial_p \alpha_k^*$ and $Q = \alpha_k \partial_q \alpha_k^*$. Equation (S13) can be further reduced to

$$\langle \partial_p \alpha_k | \partial_q \alpha_k \rangle = \partial_p \alpha_k^* \partial_q \alpha_k + \text{Im}(P)\text{Im}(Q). \quad \text{(S14)}$$

Meanwhile, using $|\alpha_k\rangle = e^{-|\alpha_k|^2/2} \sum_{n=0}^{\infty} \alpha_k^n (n!)^{-1/2} |n\rangle$, $\sum_{n=0}^{\infty} |\alpha_k|^{2n}(n!)^{-1} = e^{|\alpha_k|^2}$, and the orthonormality of Fock states, we can obtain $\langle \alpha_k | \partial_p \alpha_k \rangle = \text{Im}(P)$ and $\langle \partial_q \alpha_k | \alpha_k \rangle = \text{Im}(Q)$ from Eq. (S12). Therefore, from Eq. (S11),

$$I_{pq} = 4 \sum_{k=1}^{N} \text{Re}(\partial_p \alpha_k^* \partial_q \alpha_k). \quad \text{(S15)}$$

Let us apply Eq. (S15) to estimate Slepian-Pollak coefficients of the object source profile, $\sum_i c_i S_i(x)$, by measuring $\sum_i \gamma_i c_i S_i(x')$ (see Fig. 1b in the main text). By choosing $|\alpha_k\rangle$ as the state vector representing $k$th single Slepian-Pollak mode, we can obtain $\partial_p \alpha_k = \beta \gamma_p \delta_{pk}$ where $\delta$ and $\beta$ are the Kronecker delta and the normalisation constant, respectively. Therefore, from Eq. (S15),

$$I_{pq} = 4|\beta\gamma_p|^2 \delta_{pq}. \tag{S16}$$

Because the quantum Cramer-Rao bound for $i$th parameter estimation for a simultaneous $N$ parameters estimation, $\sigma^2_{\text{QCRB},i}$, can be represented as $\sigma^2_{\text{QCRB},i} = (I^{-1})_{ii}$, Eq. (S16) yields

$$\frac{\sigma^2_{\text{QCRB},1}}{\sigma^2_{\text{QCRB},i}} = \left|\frac{\gamma_i}{\gamma_1}\right|^2. \tag{S17}$$

Equation (S17) is the key formulation to derive the fundamental bound of the Slepian-Pollak imaging, and it is graphically shown in Supplementary Fig. S1.

Next, we quantify the achievable resolution based on the *number of degrees of freedom* argument following Ref. 9–10. Specifically, the achievable resolution can be roughly quantified by the size of field of view divided by number of reliably estimated Slepian-Pollak coefficients. The signal-to-noise ratio (i.e. actual signal variance divided by noise variance) of $i$th Slepian-Pollak coefficient estimation, SNR$_i$, becomes

$$\text{SNR}_i = \text{SNR}_1 \frac{\delta^2_{\text{QCRB},1}}{\delta^2_{\text{QCRB},i}} = \text{SNR}_1 \left|\frac{\gamma_i}{\gamma_1}\right|^2. \tag{S18}$$

Assuming scattered photons are mostly from the 1$^{\text{st}}$ Slepian-Pollak mode, SNR$_1 \approx P_{\text{tot}}$ following Poisson statistics where $P_{\text{tot}}$ is the total number of scattered photons. Assuming $\gamma_1 \approx 1$, Eq. (S18) yields

$$\text{SNR}_i = P_{\text{tot}} |\gamma_i|^2. \tag{S19}$$

Because $\gamma_i$ monotonically decreases for increasing $i$, $\text{SNR}_M \approx P_{\text{tot}}|\gamma_M|^2 \sim 1$ where $M$ is the maximum index of (and equivalently, the number of) Slepian-Pollak coefficient that can be reliably estimated, Eq. (S19) yields Eq. (1) in the main text.

**S3. Derivation of the condition for the reliable estimation of Slepian-Pollak coefficients for the utilized measurement scheme from the Cramer-Rao bound for the estimation of Slepian-Pollak coefficients.**

We first derive the $N$-by-$N$ Fisher information matrix, $J(c_1,...,c_N)$ and Cramer-Rao bound for each parameter, $\sigma^2_{\text{CRB},i}$ ($i = 1,...N$), for the simultaneous estimation of $N$ number of Slepian-Pollak coefficients of optical source profile by measuring Slepian-Pollak coefficients of optical field of radiation, with our spatial mode tomographic measurement, following[8,11].

In general, $(p, q)$ component of the Fisher information matrix for multi-parameter estimation is defined by

$$J_{pq} = -\mathbb{E}\left[\frac{\partial^2 \ln p(\mathbf{X};\boldsymbol{\theta})}{\partial \theta_p \partial \theta_q}\right] \quad (S20)$$

where $\mathbb{E}[\bullet]$, $p(\mathbf{X};\boldsymbol{\theta})$, $\mathbf{X}$, $\boldsymbol{\theta}$, and $\theta_r$ ($r = p, q$) are the expectation value of $[\bullet]$, a joint probability density function of $\mathbf{X}$, a random vector representing the measured data, a vector of parameters to be estimated, and $r$th parameter, respectively[11]. In our spatial mode tomography, $\mathbf{X}$ is a $N'$-by-1 complex random vector where $\mathbb{E}(X_k) = c_k$, and $k = 1, ..., N'$ (in general, $N' \geq N$).

Now, we find the joint probability distribution function for the $\mathbf{X}$. The complex random variable $X_k$ is experimentally obtained by following Eq. (3) in the main text,

$$X_k = \frac{1}{4}\sum_{v=1}^{4}\left[\frac{i^v X^{(k)}_{\text{a},v}}{\sqrt{I_\text{a}}} + e^{i\phi_{\text{ba}}}\frac{i^v X^{(k)}_{\text{b},v}}{\sqrt{I_\text{b}}}\right] \quad (S21)$$

where $X_{u,v}^{(k)}$ ($u$ = a, b) are the random variables of measured raw data – i.e. photon number per frame (= per repetition) at a pixel with the respective DMD mask – with the expectation value of $I_{u,v}^{(k)}$, and $I_{u,v}^{(k)} = \left| E_u^{(\text{ref})} + (-i)^v E_u^{(k)} \right|^2$. Because $E_u^{(\text{ref})}$ is strong enough for our measurement, we can assume a normal distribution for $X_{u,v}^{(k)}$ where the variance is the shot noise of the reference mode, $I_u$ [12]. Because $X_{u,v}^{(k)}$ are independent for different $u$ and $v$, and assuming $\angle\left[E_a^{(\text{ref})}\right] = 0$, the random variable $X_a^{(k)} \left(= \sum_{v=1}^{4} i^v X_{a,v}^{(k)} / \sqrt{I_a}\right)$ follows a complex normal distribution with a variance of 4 and an expectation value of $4E_a^{(k)}$ where $\left|E_a^{(k)}\right|^2$ is the expectation value of the photon number (per pixel per frame) that will be measured if $S_k^*$ is realized at the right-half of the DMD mask without anything at the left-half (see Supplementary Fig. S3). Similarly, assuming $\angle\left[e^{i\phi_{ba}} E_b^{(\text{ref})}\right] = 0$, $X_b^{(k)} \left(= e^{i\phi_{ba}} \sum_{v=1}^{4} i^v X_{b,v}^{(k)} / \sqrt{I_b}\right)$ follows a complex normal distribution with a variance of 4 and an expectation value of $4E_b^{(k)}$ where $\left|E_b^{(k)}\right|^2$ is the expectation value of the photon number (per pixel per frame) that will be measured if $S_k^*$ is realised at the left-half of the DMD mask without anything at the right-half (see Supplementary Fig. S3). Therefore, $X_k \left(= X_a^{(k)} + X_b^{(k)}\right)$ follows a complex normal distribution with a variance of 2 and an expectation value of $E^{(k)}$ ($= E_a^{(k)} + E_b^{(k)}$) where $\left|E^{(k)}\right|^2$ is the expectation value of the photon number (per pixel per frame) that will be measured if $S_k^*$ is realised at the DMD mask (see Supplementary Fig. S3). Then, the joint probability distribution for a random vector **X** becomes

$$p(\mathbf{X}, \boldsymbol{\theta}) = \prod_{k=1}^{N'} \frac{1}{2\pi} \exp\left[-\frac{\left|X_k - E^{(k)}\right|^2}{2}\right]. \tag{S22}$$

For a repetitive measurement case, more specifically, if the total number of repeated measurements for $k$th tomography is $R_{kk}$ (for all 8 sub-measurements for $k$th tomography), the joint probability distribution for the averaged random vector becomes

$$p_{\text{rep}}(\mathbf{X},\boldsymbol{\theta}) = \prod_{k=1}^{N'} \frac{1}{2\pi} \exp\left[-\frac{\left|X_k - E^{(k)}\right|^2}{2R_{kk}}\right]. \quad (S23)$$

Using Eq. (S20) and Eq. (S22), $(p, q)$ component of the Fisher information matrix for a measurement without a repetition becomes

$$J_{pq} = \frac{1}{2}\mathbb{E}\left[\sum_{k=1}^{N'} \frac{\partial^2\left[\left(X_k - E^{(k)}\right)\left(X_k^* - E^{(k)*}\right)\right]}{\partial\theta_p \partial\theta_q}\right]$$

$$= \frac{1}{2}\mathbb{E}\left[\sum_{k=1}^{N'} \partial_q\left[-\partial_p E^{(k)}\left(X_k^* - E^{(k)*}\right) - \left(X_k - E^{(k)}\right)\partial_p E^{(k)*}\right]\right]$$

$$= \frac{1}{2}\mathbb{E}\left[\sum_{k=1}^{N'}\left[-\partial_q\partial_p E^{(k)}\left(X_k^* - E^{(k)*}\right) + \partial_p E^{(k)}\partial_q E^{(k)*} + \right.\right.$$

$$\left.\left. \partial_q E^{(k)}\partial_p E^{(k)*} - \left(X_k - E^{(k)}\right)\partial_q\partial_p E^{(k)*}\right]\right]$$

$$= \sum_{k=1}^{N'} \text{Re}\left[\partial_p E^{(k)} \partial_q E^{(k)*}\right] \quad (S24)$$

where $\partial_r[\bullet] = \partial[\bullet]/\partial c_r$. Similarly, using Eq. (S20) and Eq. (S23), the $(p, q)$ component of the Fisher information matrix for a repetitive measurement becomes,

$$J_{\text{rep},pq} = \sum_{k=1}^{N'} R_{kk} \text{Re}\left[\partial_p E^{(k)} \partial_q E^{(k)*}\right]. \quad (S25)$$

We consider a $N'$-by-$N$ transfer matrix of the optical system, $T$, such that

$$\mathbf{E} = \beta T \mathbf{c} \tag{S26}$$

where $\mathbf{E} = \left(E^{(1)}, E^{(1)}, \ldots, E^{(N')}\right)^T$, $\beta$, and $\mathbf{c} = \left(c_1, c_2, \ldots, c_N\right)^T$ are the vector of expected values of tomographic measurements, the normalisation constant, and the vector of the Slepian-Pollak coefficient of an object under specific illumination, respectively, and $T$ in the superscript denotes the transpose. Then, $E^{(k)} = \beta \sum_{r=1}^{N} T_{kr} c_r$, which leads to $\partial_r E^{(k)} = \beta T_{kr}$ due to the orthogonality of Slepian-Pollak functions. Then, Eq. (S24) reduces to $J_{pq} = |\beta|^2 \sum_{k=1}^{N'} \text{Re}\left[T_{kq}^* T_{kp}\right] = |\beta|^2 \text{Re}\left[\sum_{k=1}^{N'} (T^H)_{qk} T_{kp}\right] = |\beta|^2 \text{Re}\left[(T^H T)_{qp}\right] = |\beta|^2 \text{Re}\left[(T^H T)_{pq}\right]$,

which leads to $J = |\beta|^2 \text{Re}\left[T^H T\right]$. Similarly, Eq. (S25) reduces to $J_{\text{rep},pq} = |\beta|^2 \text{Re}\left[\sum_{k=1}^{N'} (T^H)_{qk} R_{kk} T_{kp}\right] = |\beta|^2 \text{Re}\left[(T^H R T)_{qp}\right] = |\beta|^2 \text{Re}\left[(T^H R T)_{pq}\right]$ where $R$ is the $N'$-by-$N'$ repetition matrix whose $k$th diagonal element is $R_{kk}$, which leads to $J_{\text{rep}} = |\beta|^2 \text{Re}\left[T^H R T\right]$. Therefore, the accuracy of the estimation for $i$th Slepian-Pollak coefficient with the repetitive measurement where the numbers of repetitions are characterised by $R$ ($\sigma_{\text{CRB},R,i}^{-2}$), normalised by the accuracy of the estimation of $r$th Slepian-Pollak coefficient without repetition ($\sigma_{\text{CRB},1,r}^{-2}$), becomes

$$\frac{\sigma_{\text{CRB},1,r}^2}{\sigma_{\text{CRB},R,i}^2} = \frac{(J^{-1})_{rr}}{(J_{\text{rep}}^{-1})_{ii}} = \frac{[\{\text{Re}(T^H T)\}^{-1}]_{rr}}{[\{\text{Re}(T^H R T)\}^{-1}]_{ii}} \tag{S27}$$

where $r$ is the index of the reference Slepian-Pollak mode. Equation (S27) is the key formulation to derive Eq. (2) in the main text, which can be considered a realistic version of Eq. (S17).

Next, the signal-to-noise ratio (i.e. actual signal variance divided by noise variance) of $i$th Slepian-Pollak coefficient estimation with a repetition specified by $R$, $\text{SNR}_{R,i}$, becomes,

$$\text{SNR}_{R,i} = \text{SNR}_{1,r} \frac{\sigma^2_{\text{CRB},1,r}}{\sigma^2_{\text{CRB},R,i}} = \text{SNR}_{1,r} \frac{[\{\text{Re}(T^H T)\}^{-1}]_{rr}}{[\{\text{Re}(T^H RT)\}^{-1}]_{ii}} \quad (S28)$$

where $\text{SNR}_{1,r}$ is the signal-to-noise ratio of $r$th Slepian-Pollak coefficient without repetition. Assuming detected photons are mainly from the reference Slepian-Pollak mode, $\text{SNR}_{1,r} \approx P_{\text{sin}}$, where $P_{\text{sin}}$ is the measured photon number for a single tomography. Then, the condition $\text{SNR}_{R,i} \geq 1$ yields Eq. (2) in the main text.

**S4. Overall point spread function.**

Because $|\gamma_i/\gamma_1|^2$ converges to zero for larger $i$, the source amplitude of an object under a TIRS illumination can be well approximated to a truncated Slepian-Pollak series as $\mathbf{c}_o = F\mathbf{a}_o$ where $\mathbf{c}_o$, $F$, and $\mathbf{a}_o$ are an $N$-by-1 vector of Slepian-Pollak coefficients, an $N$-by-$P$ generalised Fourier transform matrix that transforms a vector from the position basis to truncated Slepian-Pollak basis, and a $P$-by-1 vector of source amplitude represented in $P$ number of discrete position basis (i.e. a vector representation of the source profile as a function of $x$ in Fig. 1b in the main text). We can expand it to represent all amplitudes with four different illumination directions (for two-dimensional imaging) for an object as $\mathbf{c}'_o = F'\mathbf{a}'_o$ where $\mathbf{a}'_o = \left(\mathbf{a}^T_{o,1}, \mathbf{a}^T_{o,2}, \mathbf{a}^T_{o,3}, \mathbf{a}^T_{o,4}\right)^T$, $F' = \text{diag}(F, F, F, F)$, $\mathbf{c}'_o = \left(\mathbf{c}^T_{o,1}, \mathbf{c}^T_{o,2}, \mathbf{c}^T_{o,3}, \mathbf{c}^T_{o,4}\right)^T$, $T$ in the superscript denotes the transpose, and $\mathbf{a}_{o,l}$ ($\mathbf{c}_{o,l}$) is the $\mathbf{a}_o$ ($\mathbf{c}_o$) for $l$th TIRS illumination. The $4N$-by-1 vector of measured Slepian-Pollak coefficients becomes $\mathbf{c}'_m = T'\mathbf{c}'_o$ where $T' = \text{diag}(T, T, T, T)$ and $T$ is the $N$-by-$N$ transfer matrix accounting for the propagation of Slepian-Pollak modes and distortion of the optical setup ($N'=N$ in our experiment). From Methods section 3, a vector linear minimum mean square error estimator for the Slepian-Pollak coefficients of the object for all illumination directions becomes $\mathbf{c}'_e = W'\mathbf{c}'_m$ where $W' = \text{diag}[W, W, W, W]$ and $W$ is the vector linear least mean squares filter matrix. The $4P$-by-1 vector of the estimated amplitude of the object in the position basis becomes $\mathbf{a}'_e = F'_{\text{inv}}\mathbf{c}'_e$ where $F'_{\text{inv}} = \text{diag}(F_{\text{inv}}, F_{\text{inv}}, F_{\text{inv}}, F_{\text{inv}})$ and $F_{\text{inv}}$ is the $P$-by-$N$ generalised inverse Fourier transform matrix that transforms a vector from the truncated

Slepian-Pollak basis to the position basis. The reconstructed object is the coherent sum of estimated amplitudes with phase corrections, which becomes $\mathbf{a}_r = V'\mathbf{a}'_e$ where $P$-by-$4P$ matrix $V' = (V_1, V_2, V_3, V_4)$, and $V_l$ is the diagonal matrix with the phase correction factor for each position for $l$th TIRS illumination derived from $V_l(\mathbf{r}) = e^{-i(\mathbf{k}_{\text{TIR},l} \cdot \mathbf{r} + \phi_l)}$ with global phase correction factor, $\phi_l = \angle \left[ \int_{|\mathbf{r}| \leq R_O} \left( \sum_{k=1}^{N} c_{e,l}(k) S_k(\mathbf{r}) \right) d\mathbf{r} \right]$ (see Eq. (4) in the main text). Then, the reconstructed amplitude of the object (e.g. polarizability) and the original amplitude of the object can be related using a linear matrix equation,

$$\mathbf{a}_r = [V'F'_{\text{inv}}W'T'F']\mathbf{a}'_o. \tag{S29}$$

We note that Eq. (S29) can be reduced to $\mathbf{a}_r = [F_{\text{inv}}WTF]\mathbf{a}_o$ for a single illumination, which can be further reduced to $\mathbf{a}_r = [F_{\text{inv}}F]\mathbf{a}_o$ for an ideal filter, which is equivalent to the definition of the theoretical point spread function of the Slepian-Pollak imaging described in Ref. 13. Equation (S29) implies that our imaging process is a linear process, more specifically, coherent linear imaging because the amplitudes are in a linear relationship (whereas intensities are in the linear relationship in incoherent imaging). The amplitude impulse response (point spread function) becomes,

$$\mathbf{p}_r = [V'F'_{\text{inv}}W'T'F']\mathbf{p}'_o \tag{S30}$$

where $\mathbf{p}'_o = \left(\mathbf{p}_o^T, \mathbf{p}_o^T, \mathbf{p}_o^T, \mathbf{p}_o^T\right)^T$ and $\mathbf{p}_o$ is a vector form of the Dirac delta function on a position basis. As a conventional coherent linear imaging, the amplitude impulse response is suitable for characterising the Slepian-Pollak imaging because Eq. (S29) with arbitrary $\mathbf{a}'_o$ can be constructed with a superposition of Eq. (S30). We note that $\mathbf{p}_r$ is slightly spatially variant, i.e. slightly different response for $\mathbf{p}_o$ with different impulse positions, mainly because Slepian-Pollak functions form a quasi-uniform basis (spatially variant, but the variation is not significant) [14]. Supplementary Fig. S9 shows calculated and experimentally obtained $P(\mathbf{r})$, which is $\mathbf{p}_r$ represented in a cartesian coordinate.

# Supplementary Figures

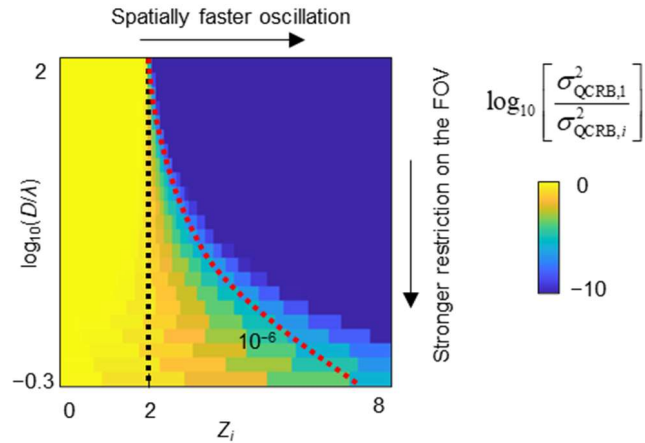

**Figure S1. Maximum attainable accuracy for the estimation of Slepian-Pollak coefficients.** Quantum Cramer-Rao bound for the estimation of Slepian-Pollak coefficients as a function of the size of the field of view. The normalised estimation accuracy, Eq. (S17), depends on both the average number of zeros per wavelength, $Z_i$ (horizontal axis), of $i$th Slepian-Pollak functions within the field of view and on the ratio $D/\lambda$ (vertical axis), where $D$ is the physical size of the field of view. The parameter $Z_i$ represents how rapidly the $i$th Slepian-Pollak function oscillates in the field of view.

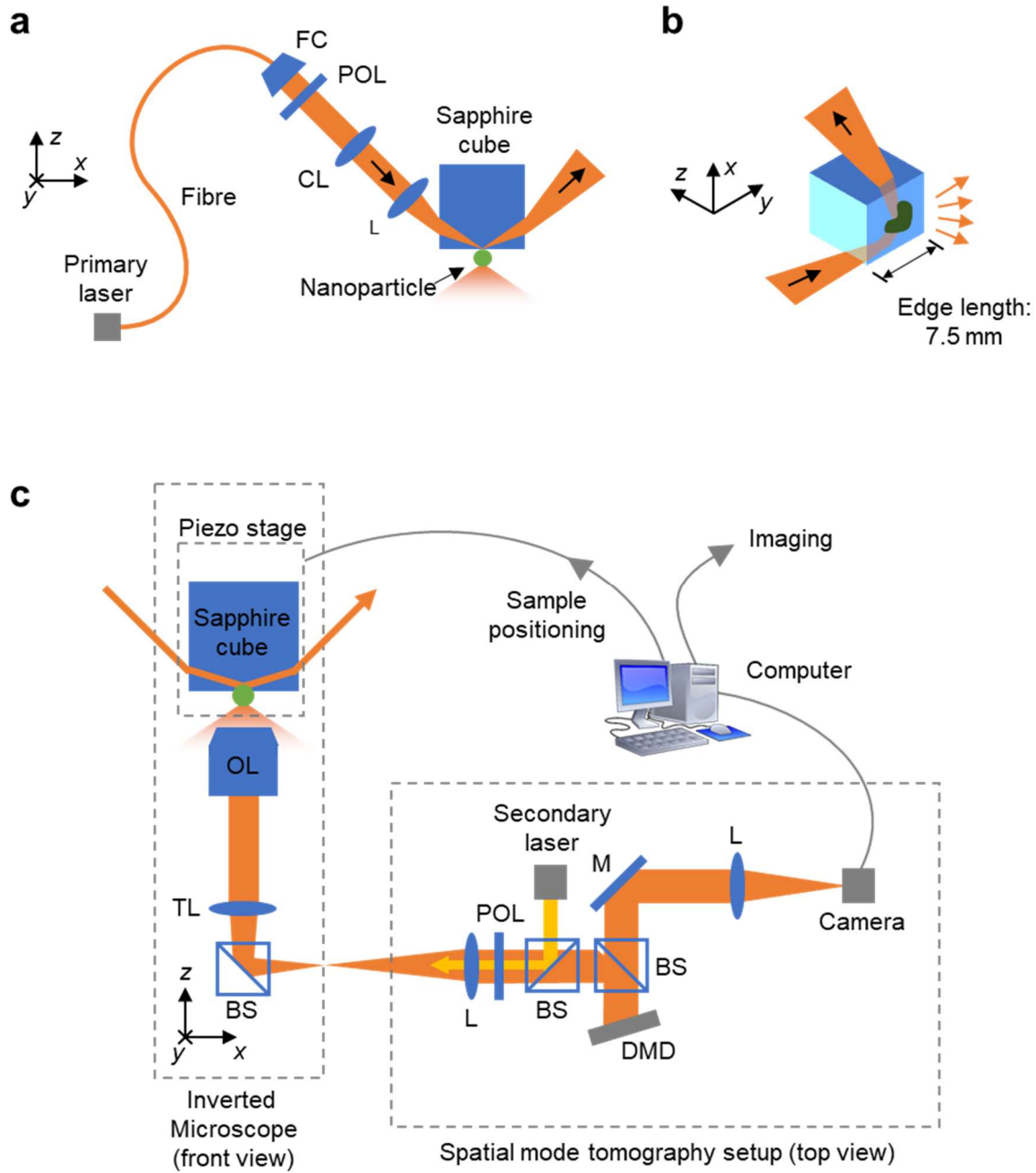

**Figure S2. Optical setup.** (a) Total internal reflection (TIR) illumination configuration realised by a sapphire cube. (b) An isometric view of sapphire cube with TIR illumination. (c) Entire optical setup. The sample (sapphire cube with nanoparticle) is placed on the inverted microscope. The scattered light is analysed with the sequential spatial mode tomography. FC: fibre collimator, POL: polarizer, CL: cylindrical lens, L: lens, OL: objective lens, TL: tube lens, BS: beam splitter, DMD: digital micromirror device, M: mirror.

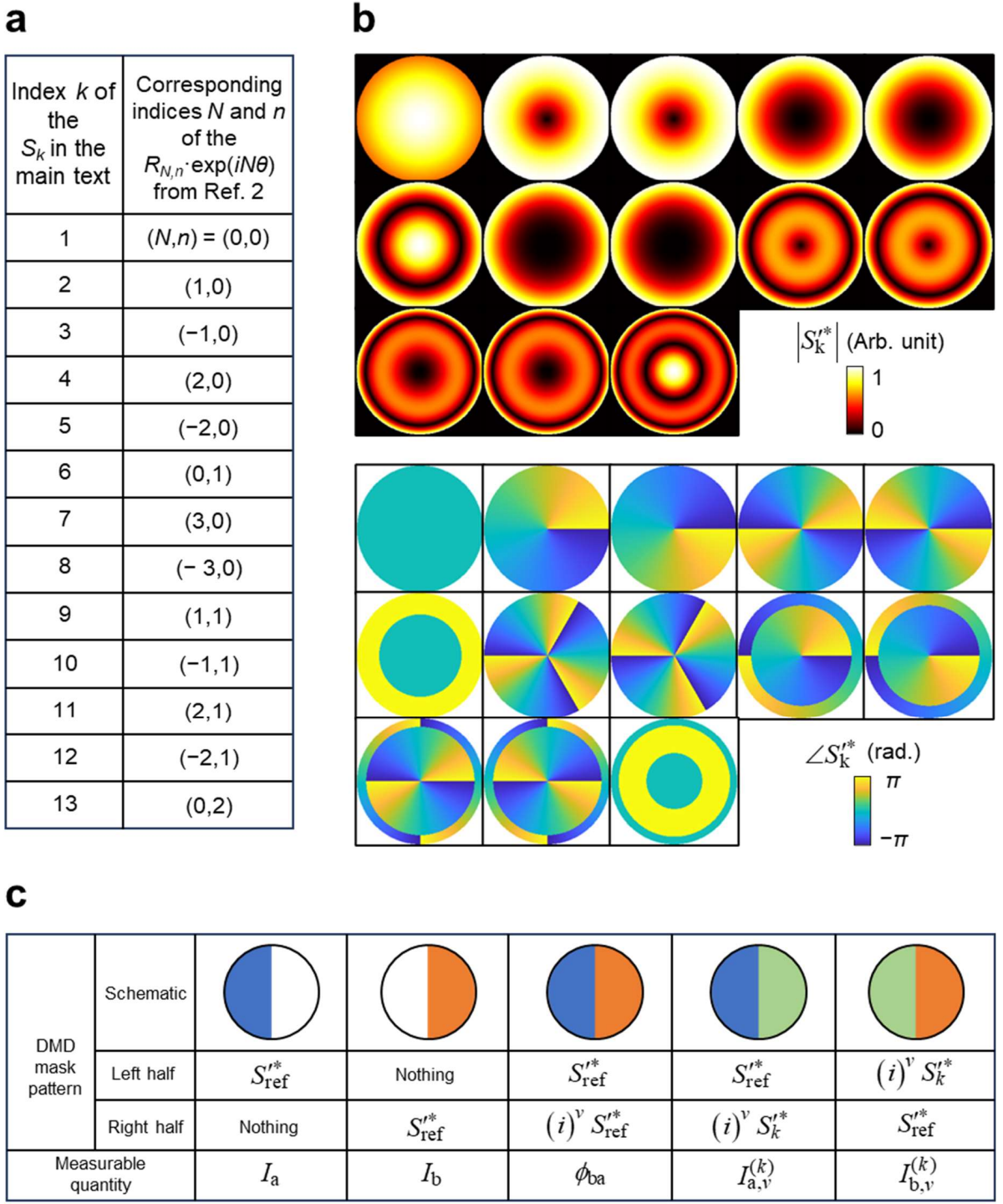

**Figure S3. Mask design for two-dimensional Slepian-Pollak imaging.** (a) Index $k$ is compared with the notation in Ref. 2. (b) Complex conjugated and renormalized (maximum modulus is set to unity) Slepian-Pollak functions, $S_k'^*$, for masks. (c) Realized mask patterns and corresponding measurable quantities are shown.

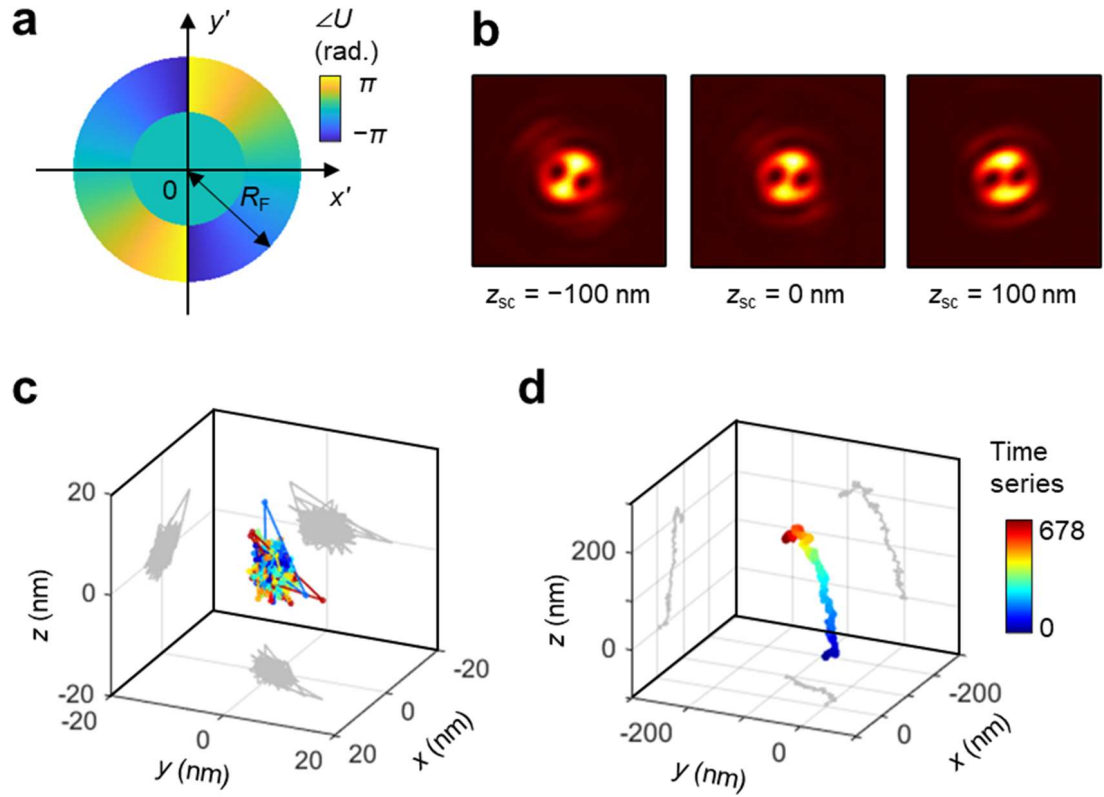

**Figure S4. Real-time drift correction.** (a) Utilized mask for simultaneous three-dimensional position estimation. Only the realized phase pattern is shown. The modulus of the mask is unity for all the mask area. (b) An example of modified image. Unlike focused spot-like image, significant change of its shape along the longitudinal displacement provides good enough estimation of the longitudinal position of nanoparticle. (c) An example showing the three-dimensional position during the entire measurement. The standard deviations for all three coordinates are less than 3 nm. (d) The amount of three-dimensional drift for the case of (c), which is corrected by the piezo stage.

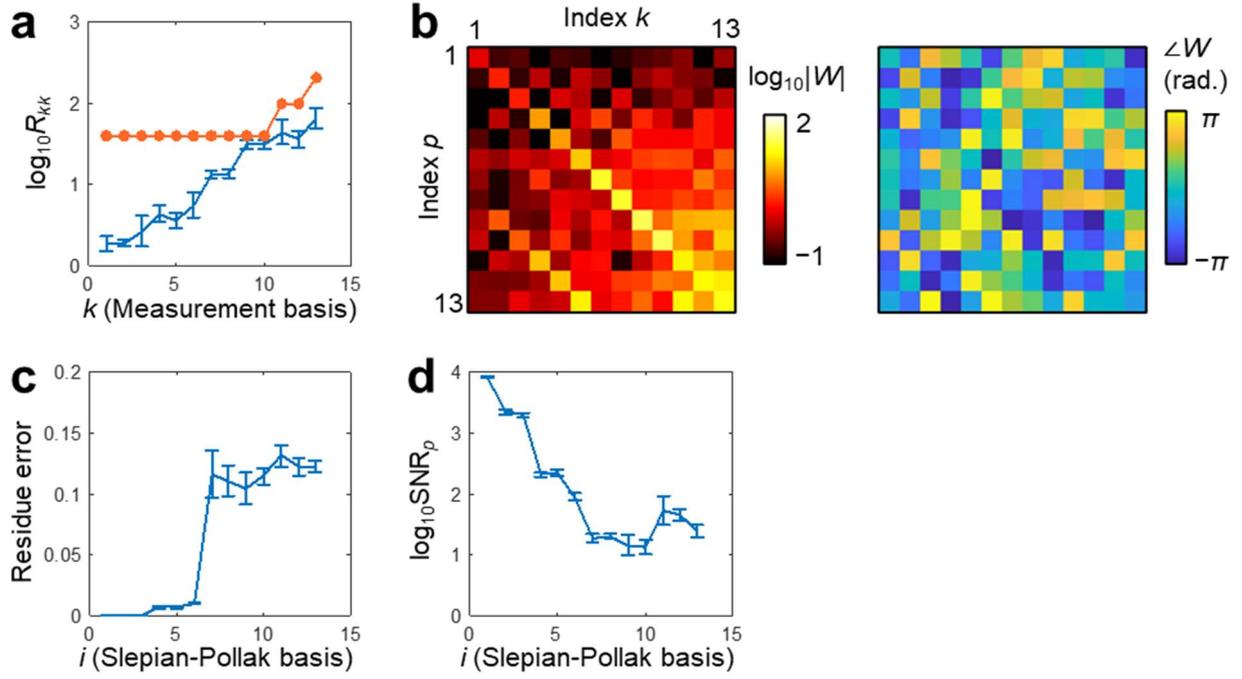

**Figure S5. Characterization of the optical setup.** (a) Required numbers of repetitions (blue) and realised numbers of repetitions (orange). We set $\varepsilon=0.1$ in Eq. (7) in the main text. (b) Experimentally constructed vector linear least mean squares (LMS) filter. (c) Average residue error between measured coefficients after filtering and the actual coefficients for the known sample (i.e. a small dot). (d) Expected signal-to-noise ratio.

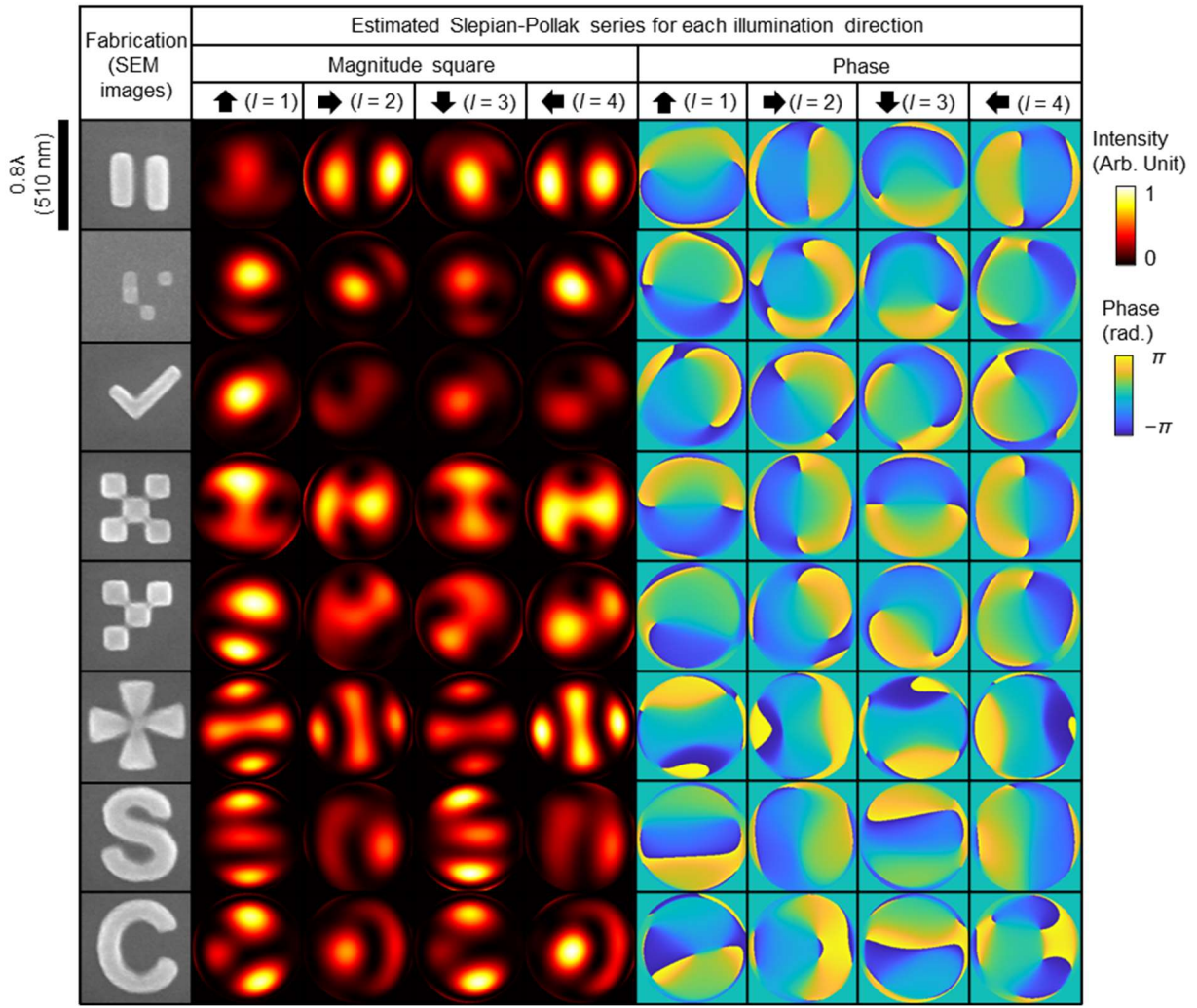

**Figure S6. Estimated Slepian-Pollak coefficients for each object for each illumination direction for two-dimensional Slepian-Pollak imaging.** An illumination direction is shown as an arrow (parallel to in-plane component of average wavevector of the incident beam).

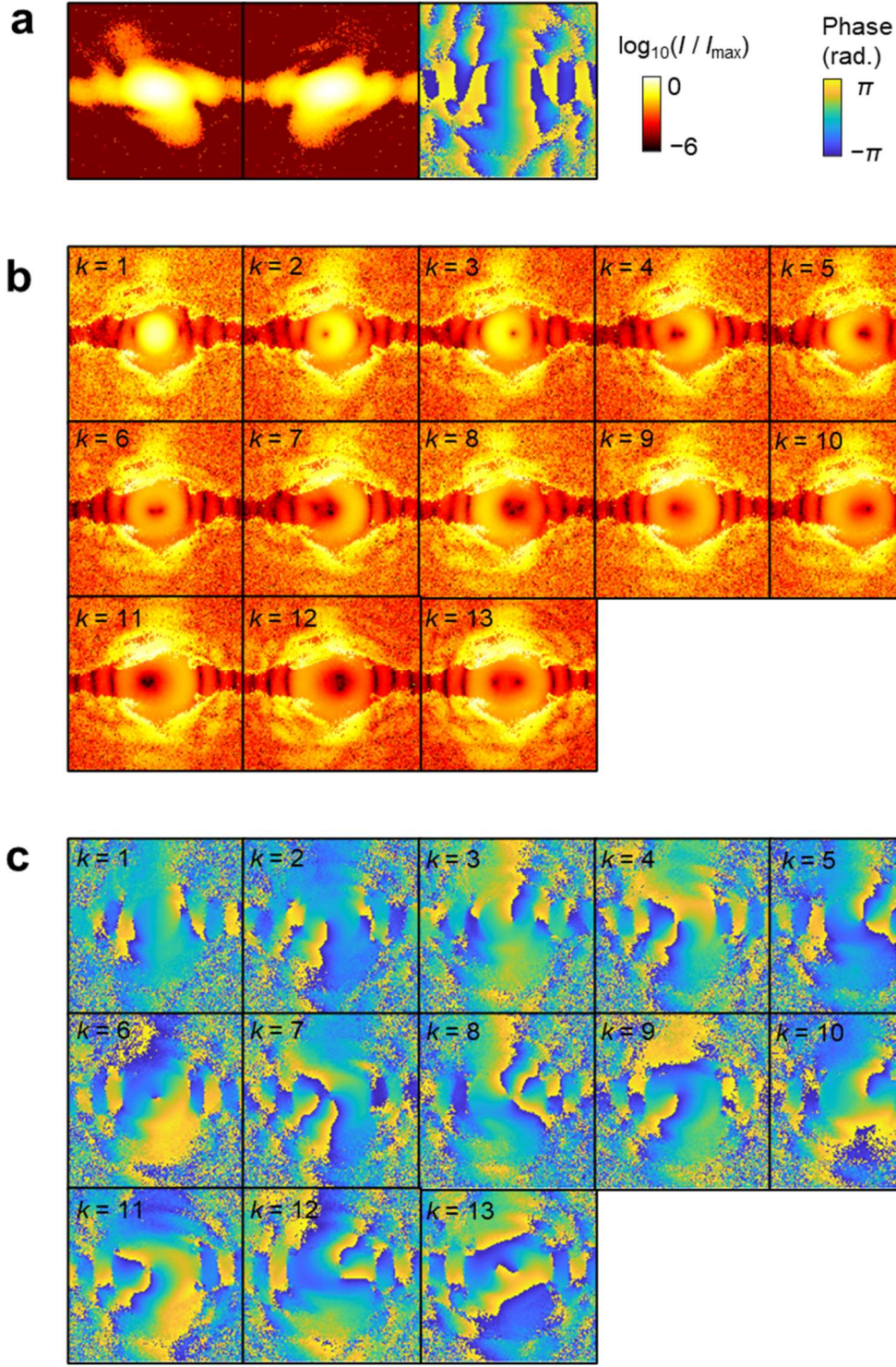

**Figure S7. A representative example ($l = 1$ case for the first object in Supplementary Fig. S6, two bars) of raw measured quantities at camera sensor pixel array, for two-dimensional Slepian-Pollak imaging.** (a) $I_a$, $I_b$, and $\phi_{ba}$, from left to right. (b) Modulus of a realization of $X_k$ (Eq. (S21)) for all $k$. (c) Phase of a realization of $X_k$ (Eq. (S21)) for all $k$.

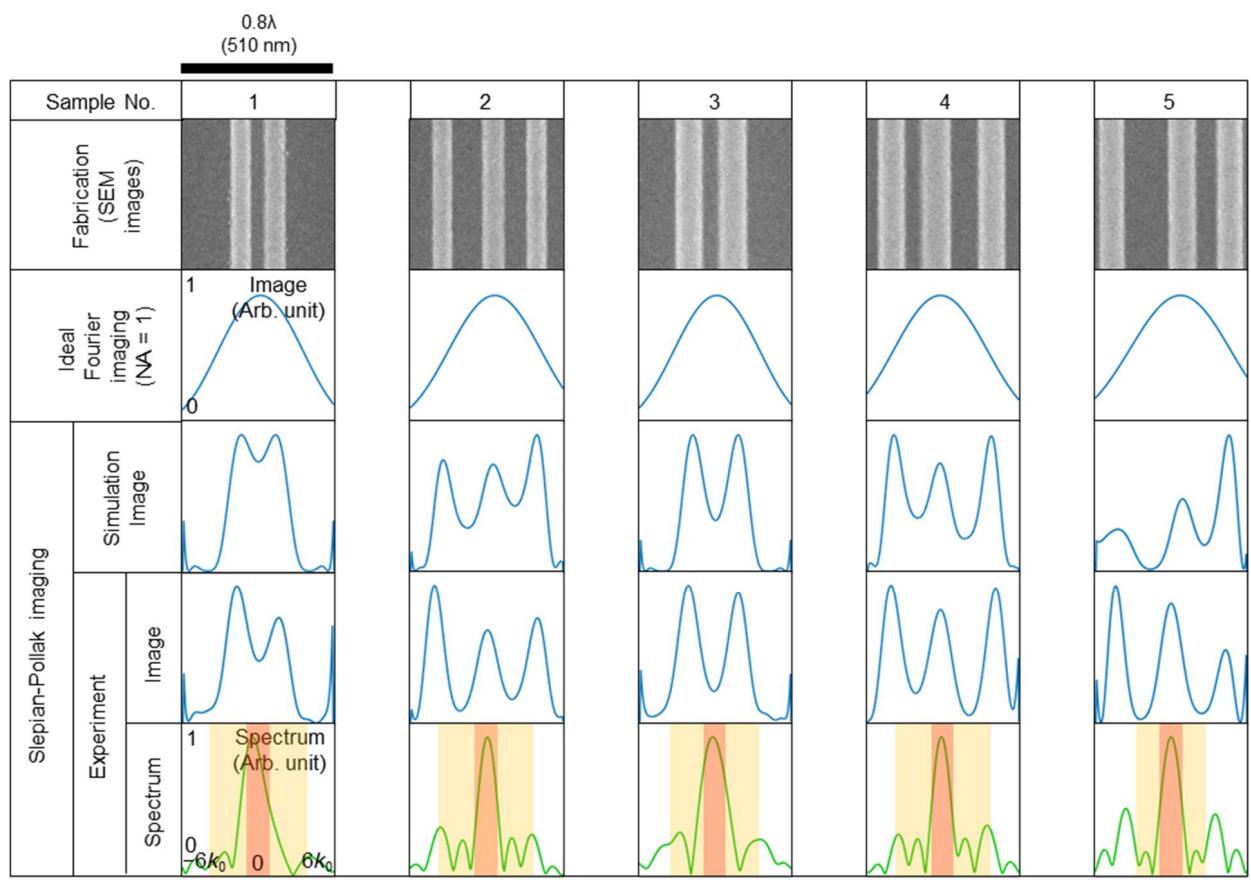

**Figure S8. Experimental demonstration of 1D Slepian-Pollak imaging.** SEM images of nanoscale lines with various numbers, widths, and gap sizes (first row from top) are compared with the calculated images for an ideal conventional Fourier imaging with numerical aperture NA=1 (second row), the simulated Slepian-Pollak images (third row) and the experimental Slepian-Pollak images (forth row), demonstrating the significant improvement in imaging performance brought about by the Slepian-Pollak imaging over the NA=1 Fourier imaging. The Spectra of the experimentally obtained images (fifth row) reveal an effective numerical aperture of 3.53.

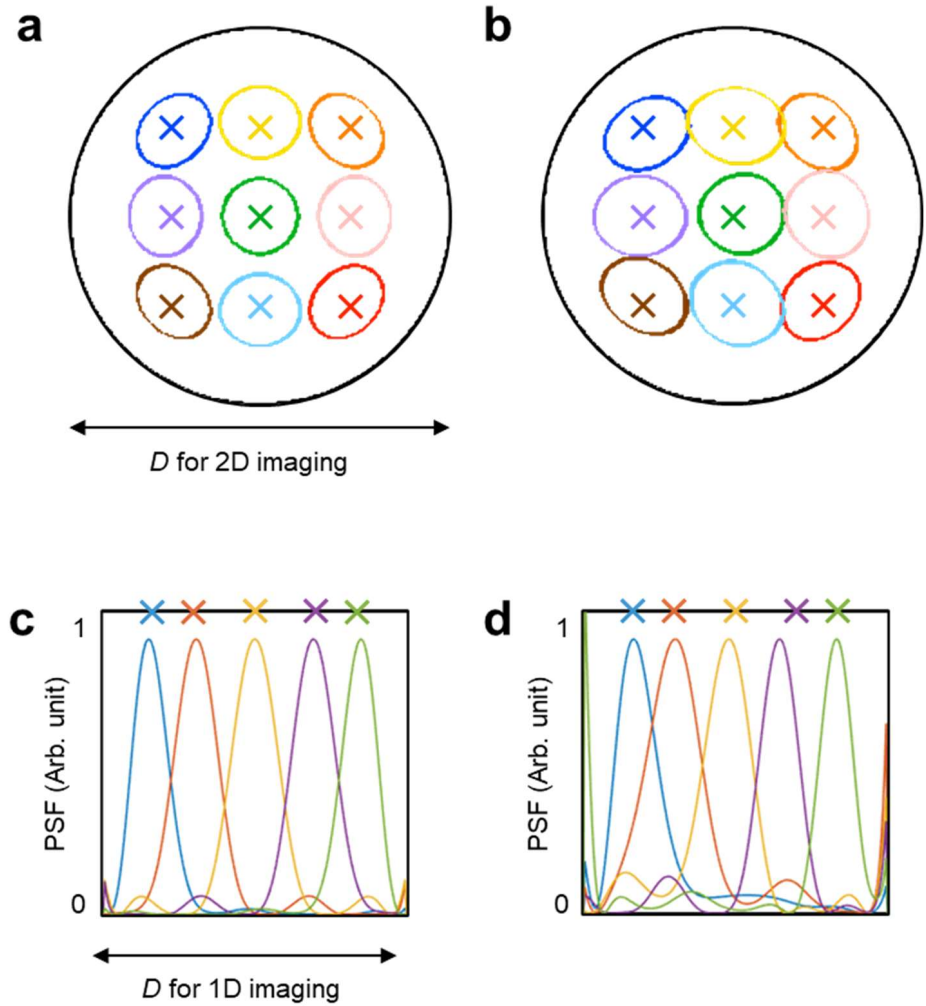

**Figure S9.** Overall point spread function (PSF) for various impulse input positions (denoted as ×). An input impulse and corresponding PSF is shown as same colour. (a, b) is for two-dimensional (half-maximum contours are shown) and (c, d) are for one-dimensional Slepian-Pollak imaging, respectively. (a, c) Calculated ideal overall PSF, i.e. modulus square of $\mathbf{p}_r = [V'F'_{inv}F']\mathbf{p}'_o$. (b, d) Experimentally obtained PSF.

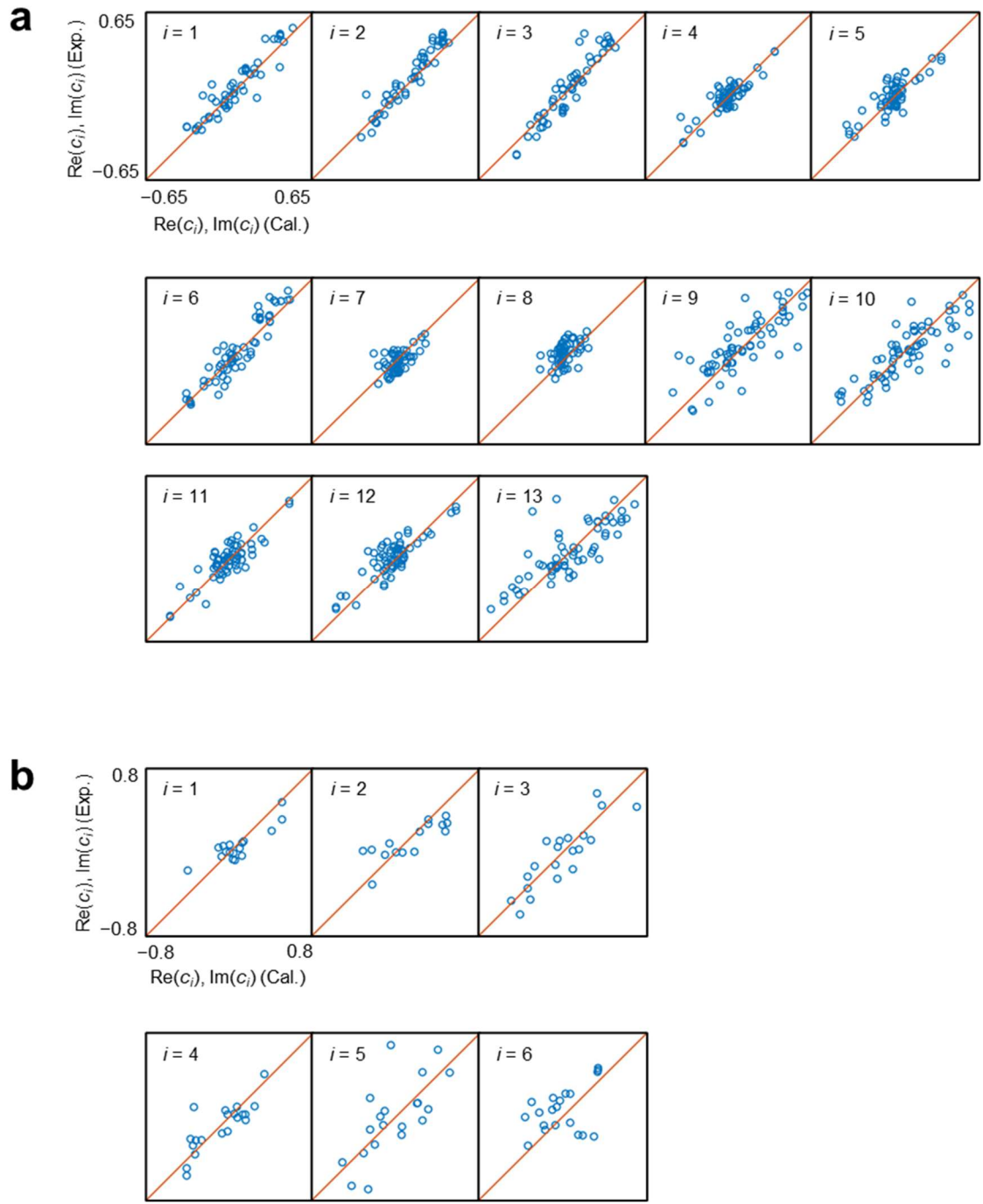

**Figure S10. Comparison between $i$th Slepian-Pollak coefficients (both real and imaginary parts) from the experiment (vertical axis) and from simulation (horizontal axis) for all $i$.** (a) For 2D imaging case. (b) For 1D imaging case.

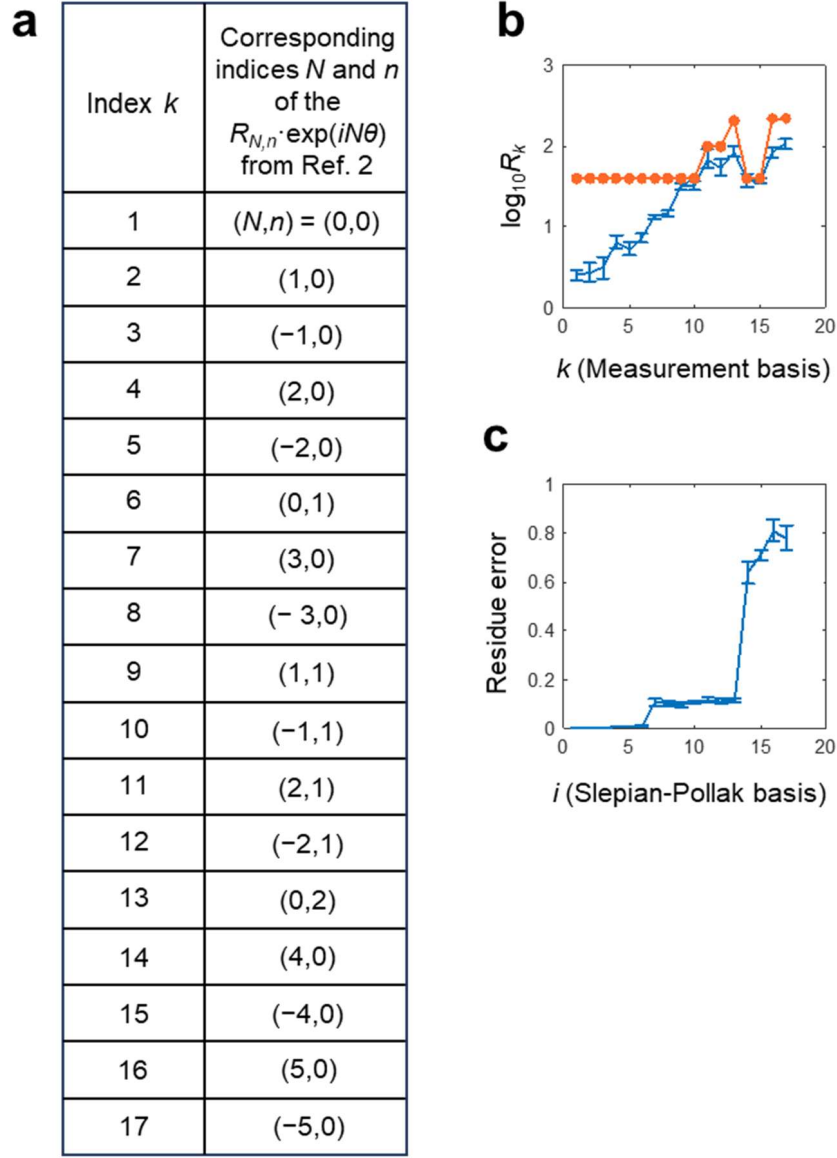

**Figure S11. Poor filter performance when 17 coefficients are measured and estimated for two-dimensional Slepian-Pollak imaging.** (a) Index $k$ is compared with the notation in Ref. 2. The same designation is used as Supplementary Fig. S3a up to $k = 13$, and additional four components are appended. (b) Required numbers of repetitions (blue) and realised numbers of repetitions (orange). Required numbers of repetitions are fulfilled. (c) Average residue error between measured coefficients after filtering and the actual coefficients for the known sample (i.e. a small dot). The significant error can be confirmed for the additional components.